\documentclass[prd,twocolumn,superscriptaddress,altaffilletter,showpacs,nofootinbib]{revtex4}
\usepackage[dvips]{graphicx}
\usepackage{amsmath}
\usepackage{graphicx,epsfig}
\newcommand{\be}{\begin{equation}}
\newcommand{\ee}{\end{equation}}
\newcommand{\bea}{\begin{eqnarray}}
\newcommand{\eea}{\end{eqnarray}}
\newcommand{\der}{\partial}

\begin{document}

\title{Cubic Derivative Interactions and Asymptotic Dynamics of the Galileon Vacuum}

\author{Roberto De Arcia}\email{dearcia@ifm.umich.mx}\affiliation{Instituto de F\'isica y Matem\'aticas, Universidad Michoacana de San Nicol\'as de Hidalgo, Edificio C-3, Ciudad Universitaria, CP. 58040 Morelia, Michoac\'an, M\'exico.}

\author{Tame Gonzalez}\email{tamegc72@gmail.com}\affiliation{Dpto. Ingenier\'ia Civil, Divisi\'on de Ingenier\'ia, Universidad de Guanajuato, Gto., M\'exico.}

\author{Genly Leon}\email{genly.leon@ucv.cl}\affiliation{Instituto de F\'isica, Pontificia Universidad Cat\'olica de Valpara\'iso, Casilla 4950, Valpara\'iso, Chile.}

\author{Ulises Nucamendi}\email{ulises@ifm.umich.mx}\affiliation{Instituto de F\'isica y Matem\'aticas, Universidad Michoacana de San Nicol\'as de Hidalgo, Edificio C-3, Ciudad Universitaria, CP. 58040 Morelia, Michoac\'an, M\'exico.}

\author{Israel Quiros}\email{iquiros@fisica.ugto.mx}\affiliation{Dpto. Ingenier\'ia Civil, Divisi\'on de Ingenier\'ia, Universidad de Guanajuato, Gto., M\'exico.}

\date{\today}

\begin{abstract} In this paper we apply the tools of the dynamical systems theory in order to uncover the whole asymptotic structure of the vacuum interactions of a galileon model with a cubic derivative interaction term. It is shown that, contrary to what occurs in the presence of background matter, the galileon interactions of vacuum appreciably modify the late-time cosmic dynamics. In particular, a local late-time attractor representing phantom behavior arises which is inevitably associated with a big rip singularity. It seems that the gravitational interactions of the background matter with the galileon screen the effects of the gravitational self-interactions of the galileon, thus erasing any potential modification of the late-time dynamics by the galileon vacuum processes. Unlike other galileon models inspired in the DGP scenario, self-accelerating solutions do not arise in this model.\end{abstract}


\maketitle


\section{Introduction}\label{sec-intro}

One of the major open questions in theoretical physics is the nature of the ``dark energy'' (DE) that is causing the observed accelerated expansion in the universe. This latter fact is supported by cosmological observations from type Ia supernovae sample from panSTARRS \cite{Riess, snia}, combined with the baryon acoustic oscillations (BAO) \cite{bao} and the cosmic microwave background (CMB) \cite{cmb}, large scale structure (LSS) \cite{lss}, weak lensing \cite{lensing}, and the integrated Sachs-Wolfe effect \cite{swe}. In the context of general relativity (GR), the cosmological constant $\Lambda$, which can be interpreted as the energy of the vacuum \cite{lambda}, provides the simplest explanation of this alien component of the cosmic budget \cite{peebles}. In the $\Lambda$CDM model, the cosmological constant accounts approximately for the 70 \% of the total energy content of the universe, meanwhile the cold dark matter (CDM) component amounts to around 25 \%. The baryonic matter and the radiation complete the cosmic inventory. This model provides the best fit to a big range of independent observations, however, there is not yet a satisfactory theoretical explanation for the very small value of $\Lambda$. Furthermore, this model suffers from a fine tuning or ``coincidence problem'' \cite{ccp}: Why is the dark matter density comparable to the vacuum energy density now, given that their time evolution is so different?

Another possibility is to consider that the DE in not a constant but evolves with time \cite{ratra}. The simplest model for an evolving dark energy are light scalar fields known as ``quintessence'', where a scalar field is postulated as the would be explanation of the observed accelerating rate of the expansion of the universe \cite{quint}. Quintessence differs from the cosmological constant in that, while the former is dynamic, that is, it changes over time, the latter remains a constant during the cosmic history. Many models of quintessence have a tracker behavior that partly solves the cosmological constant problem \cite{paul}. In these models, the quintessence field has a density which closely tracks (but is less than) the radiation density until matter-radiation equality, which triggers quintessence to start having characteristics similar to dark energy, eventually dominating the universe.

An interesting alternative explanation to the unusual cosmological dynamics at large scales, without the need to invoke a cosmological constant or a negative pressure fluid, is based on the modification of the left-hand side (LHS) of the Einstein field equations. Among the most representative models that belong in this sector are those based on $f(R)$ theories \cite{Capozziello, Hu}, $f(R, \mathcal{G})$ theories \cite{Carroll}, Brans-Dicke (BD) theories \cite{Brans}, Dvali-Gabadadze-Porrati (DGP) braneworld \cite{Dvali}, and Galileon gravity \cite{nicolis, deffayet, deffayet-rev}. The $f(R)$ gravity is a kind of generalization of Einstein's GR which can be put into the form of a scalar-tensor theory (BD theory with vanishing coupling constant $\omega=0$ to be precise). As a matter of fact $f(R)$ theories represent a family of models, each one defined by a different function of the Ricci scalar. GR is recovered after the simplest choice of the function $f(R)\propto R$. As a consequence of introducing an arbitrary function there is a lot of freedom to explain the accelerated expansion and structure formation of the Universe without adding unknown forms of dark energy or dark matter. The BD theory represents a particular case of a scalar-tensor theory, a gravitational theory in which in addition to the metric field the gravitational interaction is mediated also by a scalar field. The gravitational coupling constant $G$ is not presumed to be constant but instead $1/G$ is replaced by a point-dependent scalar field. 

The DGP model assumes the existence of a five-dimensional (5D) Minkowski space of infinite volume within which ordinary four-dimensional (4D) Minkowski space is embedded. Consequently, there is no normalizable zero-mode of the 4D graviton in the DGP brane-world. The action of this model consists of two pieces: a standard 4D Einstein-Hilbert action, which involves only the known 4D spacetime dimensions, plus the Einstein-Hilbert action of a 5D Minkowski manifold. The transition from 4D to 5D behavior is governed by a crossover scale $r_c$ \cite{roy}: gravity leaks off the 4D brane into the bulk at large scales, $r\gg r_c$, where the 5D piece of the action dominates, while on small scales, $r\ll r_c$, gravity is effectively bound to the brane and 4D dynamics is recovered to a good approximation, as the 4D piece of the action dominates. 

In spite of the very attractive features of the DGP model, it is ruled out by observations in comparison with the $\Lambda$CDM model \cite{roy-majeroto}. Besides, in addition to the insurmountable problems posed by the cosmological observations, there is a problem of theoretical consistency associated with the existence of a ghost in the late-time asymptotic de Sitter solution. This is a ghost mode in the scalar sector of the effective Brans-Dicke theory that approximates the DGP on cosmological subhorizon scales, which is more serious than the ghost in a phantom scalar field \cite{roy}.

Finally we mention the so called galileon models which will be the focus of the present investigation. Inspired by the DGP model, in \cite{nicolis} the authors derived the five Lagrangians that lead to field equations invariant under the Galilean symmetry $\der_\mu\phi\rightarrow\der_\mu\phi+b_\mu$ in the Minkowski space-time. The scalar field that respects the Galilean symmetry is dubbed ``galileon''. Each of the five terms only leads to second-order differential equations, keeping the theory free from unstable spin-2 ghost degrees of freedom. If we extend the analysis in Ref. \cite{nicolis} to the curved space-time, the Lagrangians need to be promoted to the covariant forms. Deffayet et al. \cite{deffayet} derived the covariant Lagrangians ${\cal L}_i$ ($i=1,..., 5$) that keep the field equations up to second-order. These are equivalent to the ones discovered by Horndeski \cite{horndeski}. 

The most general 4-dimensional scalar-tensor theories having second-order field equations are described by the linear combinations of the following Lagrangians \cite{deffayet}:

\begin{widetext}\bea 
&&{\cal L}_2=K,\;{\cal L}_3 =-G_3(\nabla^2\phi),\;{\cal L}_4=G_4 R+G_{4,X}\left[(\nabla^2\phi)^2-2(\nabla\phi)^2\right],\nonumber\\
&&{\cal L}_5=G_5 G_{\mu\nu}(\nabla^{\mu}\nabla^\nu\phi)-\frac{1}{6}G_{5,X}\left[(\nabla^2\phi)^3-3(\nabla^2\phi)(\nabla_\mu\nabla_\nu\phi)(\nabla^\mu\nabla^\nu\phi)+2(\nabla^\mu\nabla_\alpha\phi)(\nabla^\alpha\nabla_\beta\phi)(\nabla^\beta\nabla_\mu\phi)\right],\eea\end{widetext} where $(\der\phi)^2:=g^{\mu\nu}\nabla_\mu\phi\nabla_\nu\phi$, $\nabla^2\phi:=g^{\mu\nu}\nabla_\mu\nabla_\nu\phi$, $K=K(\phi,X)$ and $G_i=G_i(\phi,X)$ ($i=3,4,5$), are functions of the scalar field $\phi$ and its kinetic energy density $X=-(\der\phi)^2/2$, while $G_{i,\phi}$ and $G_{i,X}$, represent the derivatives of the functions $G_i$ with respect to $\phi$ and $X$, respectively.

In order to perform the dynamical analysis of the generalized Galileon cosmology we need to focus in specific models. One class of Galileon scenarios of particular interest for cosmology has the following choice ($G_5=0$):

\bea K=X-V(\phi),\;G_3=gX,\;G_4=\frac{1}{16\pi G_N},\label{model}\eea where $g=g(\phi)$ is a coupling function and $G_N$ is the gravitational coupling (Newton's) constant. The resulting action reads:

\bea S=\int d^4x\frac{\sqrt{-g}}{2}\left\{R-\left[1+g\nabla^2\phi\right](\der\phi)^2-2V(\phi)\right\},\label{action}\eea where, $G_{\mu\nu}=R_{\mu\nu}-g_{\mu \nu}R/2$, is the Einstein's tensor and $V=V(\phi)$, is the self-interaction potential of the galileon. The matter action piece $S_m=\int d^4x\sqrt{-g}\,{\cal L}_m$, where ${\cal L}_m$ stands for the matter Lagrangian, has been omitted for simplicity but, if desired, it may be added.
 
The cubic derivative interaction term $\propto f(\phi)\nabla^2\phi(\der\phi)^2$ has been investigated before in the context of the Brans-Dicke theory in \cite{silva_koyama} (see also \cite{khoury}). In the mentioned theory this term is the responsible for the approximate recovering of general relativity on small scales and at early times. This is the unique form of interactions at cubic order yielding second order motion equation for the galileon field. This is essential for physical theories since the extra degrees of freedom associated with the higher-derivatives usually generate instabilities.

Given the extremely complex form of the generalized galileon field equations which are obtained by means of the variational principle from \eqref{action}, deriving of exact cosmological solutions is by far a mammoth task. This is where the tools of the dynamical systems theory come into scene. Although the phase space dynamics of the class of models especified by the choice \eqref{model} has been investigated in detail in Ref. \cite{genly} for a pair of choices of the coupling function $g=g(\phi)$ and of the potential $V(\phi)$, here we want to pay special attention to a particular case that was not investigated in that reference: the galileon vacuum cosmology. For the more general situation when, in addition to the galileon, the background matter is considered, in \cite{genly} it was found that there are not any new stable late-time solutions apart from those of standard quintessence. In consequence one may naively expect that the same result should hold true for the particular case when the standard matter degrees of freedom are removed. The results of this paper will show quite the contrary: there is a very interesting asymptotic dynamics in the vacuum of the generalized galileon cosmological models, which strongly departs from the asymptotic structure of standard quintessence even at late-time.

For a better understanding of the subject we shall discuss here, the paper has been organized in the following way. The basic information on the model subject of this investigation, is exposed in Sec. \ref{sec-basic}. Then, in Sec. \ref{sec-var}, we discuss on the adequate choice of the variables of the phase space. When cosmological issues are involved this is, by far, one of the most important topics of the dynamical systems study since not every choice leads to correct results: an inappropriate choice of the phase space variables may lead to the loss of one or more important equilibrium configurations. In Sec. \ref{sec-mat} the generalized galileon model with cubic derivative interaction $\propto g_0\nabla^2\phi(\der\phi)^2$ (the coupling $g(\phi)=g_0$ is a constant), and with exponential potential $V(\phi)\propto\exp(-\lambda\phi)$, is explored in the presence of a background matter fluid, as an illustration of the dynamical systems approach used in the present paper. Recall that this and more general cases have been studied before in \cite{genly}. In Sec. \ref{sec-vac} we dispense with the matter component and we focus in the apparently simpler case when the cosmic background is just vacuum. The result obtained can not be more surprising: the asymptotic cosmological dynamics of the vacuum galileon is richer than the asymptotic dynamics in the presence of background matter. In particular a late-time phantom attractor associated with a big rip singularity arises. This and other obtained results are discussed in Sec. \ref{sec-discuss}, while in Sec. \ref{sec-conclu} conclusions are given. In this paper we use the units where $8\pi G=c=h=1$.


\section{basic setup and cosmological equations}\label{sec-basic}

Here a Friedmann-Robertson-Walker (FRW) spacetime with flat spatial sections is assumed, whose line element is given by $ds^2=-dt^2+a^2(t)\delta_{ik}dx^idx^k.$ The cosmological field equations resulting from the action \eqref{action}, read:

\bea &&\;3H^2=\rho_m+\rho_\phi,\nonumber\\
&&-2\dot H=\rho_m+p_m+\rho_\phi+p_\phi,\nonumber\\
&&\left(1+2g_{,\phi}\dot\phi^2-6gH\dot\phi\right)\ddot\phi+3H\dot\phi \nonumber\nonumber\\
&&\;\;\;\;+\left(\frac{1}{2}g_{,\phi\phi}\dot\phi^2-3g\dot H-9gH^2\right)\dot\phi^2=-V_{,\phi},\nonumber\eea where, in addition to the galileon, a standard matter fluid with energy density $\rho_m$ and barotrotopic pressure $p_m$, is assumed. The energy density and the parametric pressure of the galileon field are given by

\bea &&\rho_\phi=\frac{\dot\phi^2}{2}\left(1+g_{,\phi}\dot\phi^2-6gH\dot\phi\right)+V,\nonumber\\
&&p_\phi=\frac{\dot\phi^2}{2}\left(1+g_{,\phi}\dot\phi^2+2g\ddot\phi\right)-V.\eea 

In this paper, for simplicity of the analysis, we shall focus in the constant galileon coupling case, with the exponential potential:

\bea g=g_0\;\Rightarrow\;g_{,\phi\phi}=g_{,\phi}=0,\;V(\phi)=V_0\,e^{-\lambda\phi}.\label{case}\eea Besides, for definiteness we shall assume non-negative $g_0\geq 0$, which is the more interesting choice, since for $g_0<0$, the asymptotic dynamics is not as interesting, resulting in a straightforward particular case of that of galileon cosmology with background matter. Under the above assumptions the cosmological Einstein's field equations read:

\bea &&\;3H^2=\rho_m+\rho_\phi,\nonumber\\
&&-2\dot H=\rho_m+p_m+\rho_\phi+p_\phi,\label{feqs}\eea while the motion equation of the galileon is depicted by:

\bea &&\left(1-6g_0H\dot\phi\right)\ddot\phi+3H\dot\phi\nonumber\\
&&\;\;\;\;\;\;\;\;\;\;\;\;\;\;\;\;\;\;\;\;\;\;-3g_0H^2\left(3+\frac{\dot H}{H^2}\right)\dot\phi^2=-V_{,\phi}.\label{kg-eq}\eea In the above equations:

\bea &&\rho_\phi=\frac{\dot\phi^2}{2}\left(1-6g_0H\dot\phi\right)+V,\nonumber\\
&&p_\phi=\frac{\dot\phi^2}{2}\left(1+2g_0\ddot\phi\right)-V.\label{rho-p}\eea 

Equations \eqref{feqs}, \eqref{kg-eq}, \eqref{rho-p}, are the master equations of the model which is the subject of the present investigation.


\section{The variables of the phase space}\label{sec-var}

Our aim here will be to trade the very complex system of second order equations \eqref{feqs}, \eqref{kg-eq}, \eqref{rho-p}, by a system of autonomous ordinary differential equations (ODE-s). For this purpose one has to choose adequate variables of some state space. To start with one chooses the following standard, Hubble-normalized variables of the phase space \cite{wands}:

\bea &&x_s=\frac{\dot\phi}{\sqrt{6} H},\;y_s=\frac{\sqrt V}{\sqrt{3}H}.\label{xy-var}\eea 

In terms of these variables the Friedmann equation in (\ref{feqs}) can be written as:

\bea \Omega_m=1-x_s^2-y_s^2+6\sqrt{6}\,x^3_sH^2g_0,\label{friedmann-eq}\eea where $\Omega_i:=\rho_i/3H^2$ is the dimensionless (normalized) energy density of the $i$-th matter component. As seen from Eq. \eqref{friedmann-eq}:  

\begin{enumerate}

\item One needs yet another phase space variable to account for the factor $H^2g_0$.

\item Due to the positive sign of the fourth term in the right-hand side (RHS) of Eq. (\ref{friedmann-eq}), given $x_s\geq 0$, the variables $x_s$ and $y_s$ can take arbitrary large values, while $0\leq\Omega_m\leq 1$.

\end{enumerate} 

In consequence, we introduce the following bounded new variables of the phase space:

\bea x_\pm=\frac{1}{x_s\pm1},\;y=\frac{1}{y_s+1},\;z=\frac{1}{H^2g_0+1},\label{n-var}\eea where $x_+$ is for non-negative $x_s\geq 0$ ($\dot\phi\geq 0$), while $x_-$ is for non-positive $x_s\leq 0$ ($\dot\phi\leq 0$). Besides, $0\leq x_+\leq 1$ ($-1\leq x_-\leq 0$), $0\leq y\leq 1$, and $0\leq z\leq 1$. Here we are assuming that only expanding cosmologies arise: $H\geq 0$ ($y_s\geq 0$), and that along orbits of the phase space $x_s$ does not flip sign. These assumptions are not independent of each other. Actually, at a bounce, no matter whether it is a bounce at a minimum or at a maximum size of the universe, where $\dot a=0,\;\ddot a>0$ (minimum size), or $\dot a=0,\;\ddot a<0$ (maximum size universe), since $H$ flips sign, then, necessarily $y_s\propto\sqrt{V}/H$, flips sign as well. Notice that the bounce, if present, arises at the boundary $y_s=0$ since, while $H$ flips sign $\sqrt{V}$ does not. Besides, at the bounce, simultaneously, $\dot\phi\sim H\sim 0$ and $\sqrt{V}\sim H\sim 0$ since otherwise, if assume finite $\dot\phi$ and $V$, $$x_s=\frac{\dot\phi}{\sqrt{6}H}\rightarrow\infty,\;y_s=\frac{\sqrt V}{\sqrt{3}H}\rightarrow\infty.$$

The choice of coordinates in (\ref{n-var}) is specially useful in those cases where $x_s=0$, and $y_s=0$ are invariant subspaces in the ($x_s,y_s$) -- phase space. This means that orbits originated from initial conditions, say, in the quadrant $x_s\geq 0$, $y_s\geq 0$, will entirely lay in that quadrant. I. e., the orbits will not cross none of the boundaries (it could be better to say separatrices), $x_s=0$ and $y_s=0$. As we will show below, this is, precisely, the case for the vacuum of the generalized galileon model \eqref{feqs}, \eqref{kg-eq}, \eqref{rho-p}, which is the subject of the present investigation. 

In what follows it will be useful to define the quantity:

\bea Q:=-9H^2g_0=9\left(\frac{z-1}{z}\right)<0,\label{Q-def}\eea so that, in particular, the Friedmann constraint (\ref{friedmann-eq}) can be written as:

\bea \Omega_m=1-x_s^2-y_s^2-2\sqrt\frac{2}{3}\,x^3_s Q.\label{friedmann-c}\eea


\begin{table*}\centering
\begin{tabular}{|c||c|c|c||c|c|c|c|c|c|c|c|}
\hline\hline
Crit. Point&\;\;\;$x_\pm$\;\;\;&\;\;\;$y$&\;\;\;$z$\;\;\;\;\;&Exist.&\;\;\;$q$\;\;\;&\;\;\;$\lambda_1$\;\;\;&\;\;\;$\lambda_2$\;\;\;&\;\;\;$\lambda_3$\;\;\;& Stab. & $\Omega_m$ & $\omega_\phi$ \\
\hline\hline
$P^\pm_1$&$\pm 1$ & $1$ & $0$& always & $\frac{1}{2}$ & undef. & undef. & undef. & unstable  & $1$ & undef. \\
&&&&&&&&& (num. invest.) &&\\
\hline
$P^\pm_2$&$\pm 1$ & $1$ & $1$ & '' & $\frac{1}{2}$ &$-\frac{3}{2}$ & $-3$ & $\frac{3}{2}$ & saddle & $1$ & $1$ \\
\hline
$P^\pm_3$&$\pm\frac{1}{2}$ & $1$ & $1$ & '' & $2$ & $3\mp\sqrt\frac{3}{2}\lambda$ & $3$ & $-6$ & saddle & $0$ & $1$ \\
\hline
$P^\pm_4$&$\frac{\sqrt{6}}{\lambda\pm\sqrt{6}}$ & $\frac{\sqrt{6}}{\sqrt{6-\lambda^2}+\sqrt{6}}$& $1$ & $\lambda^2< 6$ & $-1+\frac{\lambda^2}{2}$ & $-\lambda^2$ & $-3+\frac{\lambda^2}{2}$ & $-3+\lambda^2$ & stab. if $\lambda^2<3$ & $0$ & $-1+\frac{\lambda^2}{3}$ \\
& & & & & & & & & sadd. if $\lambda^2>3$ & & \\
\hline
$P^\pm_5$& $\frac{\pm 2\lambda}{2\lambda\pm\sqrt{6}}$ & $\frac{2\lambda}{2\lambda\pm\sqrt{6}}$ & $1$ & $\lambda^2>3$ & $\frac{1}{2}$ & $-3$ & $-\frac{3}{4}+\alpha$ & $-\frac{3}{4}-\alpha$ & stable point & $\frac{\lambda^2-3}{\lambda^2}$ & $0$ \\
& & & & & & & & & spiral if $\lambda^2>\frac{24}{7}$ & & \\
\hline\hline
\end{tabular}\caption{Critical points of the dynamical systems (\ref{mat-asode}), together with their main properties. We have used the following parameter definition: $\alpha:=\sqrt{-7+24/\lambda^2}$.}\label{tab-1}
\end{table*}


\section{generalized galileon cosmology with matter}\label{sec-mat}

In spite of the fact that this is a particular case of the more general situation investigated in \cite{genly}, in order to illustrate our adopted procedure which, to be honest, is clearly different from the one undertaken in the mentioned bibliographic reference, here we shall study the generalized galileon cosmology in the presence of a matter fluid with energy density $\rho_m$ and barotropic pressure $p_m$. Here, for simplicity, we set $p_m=0$, so we deal with background (pressureless) dust. As mentioned before, it will be adopted the exponential potential $V(\phi)=V_0\exp(-\lambda\phi)$, and the constant galileon coupling $g=g_0$, will be assumed \eqref{case}. 

The next step is to trade the cosmological field equations \eqref{feqs}, \eqref{kg-eq}, $\eqref{rho-p}$, by the following dynamical system given in terms of the bounded variables $x_\pm$, $y$, and $z$, which were defined in Eq. \eqref{n-var}:

\begin{widetext}
\bea x'_\pm=-\frac{x^2_\pm}{\sqrt{6}}\,\eta_\pm+x_\pm(1\mp x_\pm)\gamma_\pm,\;y'=y(1-y)\left[\sqrt\frac{3}{2}\,\lambda\left(\frac{1\mp x_\pm}{x_\pm}\right)+\gamma_\pm\right],\;z'=-2z(1-z)\gamma_\pm,\label{mat-asode}\eea\end{widetext} where, for compactness of writing, we have defined:

\begin{widetext}
\bea &&\gamma_\pm:=\left[\frac{\dot H}{H^2}\right]_\pm=-3\left\{\frac{\frac{3}{2}\,x_\pm^2\Theta_\pm(1)+2(1\mp x_\pm)^4y^2Q^2+x_\pm(1\mp x_\pm)\left[\sqrt{6}\Theta_\pm(2)-\lambda x_\pm(1\mp x_\pm)(1-y)^2\right]Q}{y^2\left[3x_\pm^4+2\sqrt{6}x_\pm^3(1\mp x_\pm)Q+2(1\mp x_\pm)^4Q^2\right]}\right\},\nonumber\\
&&\eta_\pm:=\left[\frac{\ddot\phi}{H^2}\right]_\pm=-9\left\{\frac{\sqrt{6} x_\pm^3(1\mp x_\pm)y^2-\lambda x_\pm^4(1-y)^2+(1\mp x_\pm)^2\Gamma_\pm Q}{y^2\left[3x_\pm^4+2\sqrt{6}x_\pm^3(1\mp x_\pm)Q+2(1\mp x_\pm)^4Q^2\right]}\right\}.\label{def-s}\eea\end{widetext} We have defined also the following functions:

\begin{widetext}
\bea\Theta_\pm(a):=a(1\mp x_\pm)^2y^2-x^2_\pm(1-2y),\;\Gamma_\pm:=x_\pm^2(1-y)^2-(1\mp2 x_\pm)y^2.\label{defs}\eea\end{widetext} In Eq. \eqref{mat-asode} the comma denotes the derivative $f'=H^{-1} \dot f$, while the '$\pm$' signs above account for two different branches of the dynamical system so that, as a matter of fact, one has two different dynamical systems.

\subsection{Critical points in the 3D phase space}

The critical points of the dynamical systems (\ref{mat-asode}), together with their main properties, can be found in TAB. \ref{tab-1}. This table reflects the fact that, but for the matter-dominated big bang, which is associated with the critical point $P_1^\pm$, as clearly stated in \cite{genly}, the present galileon model (perhaps others too) does not differ too much from the standard quintessence. This is particularly true for the late-time dynamics (points $P_4^\pm$ and $P_5^\pm$). Notice that the big bang solution is not a global past attractor but a local one. The remaining equilibrium points are the same found in TAB. 1 of Ref. \cite{wands}: 

\begin{enumerate}

\item The matter dominated solution $P_2^\pm$, which is associated with a saddle critical point. 

\item The stiff-matter solutions $P_3^\pm$, which are correlated with a scalar field's kinetic energy density dominated universe. In the present case these are always saddle points, while in the standard quintessence case these can be the past attractors as well. 

\item The scalar field dominated solution $P_4^\pm$, representing scaling between the scalar field's kinetic and potential energy densities. This can be either a saddle or a late-time attractor as in the quintessence model. 

\item The scalar field-matter scaling solution $P_5^\pm$. Whenever it exists it is a late-time attractor. It is either a focus or an spiral equilibrium point.

\end{enumerate} 

The above results are essentially the same obtained in \cite{genly} by means of a bit different procedure.


\begin{figure*}
\includegraphics[width=5cm]{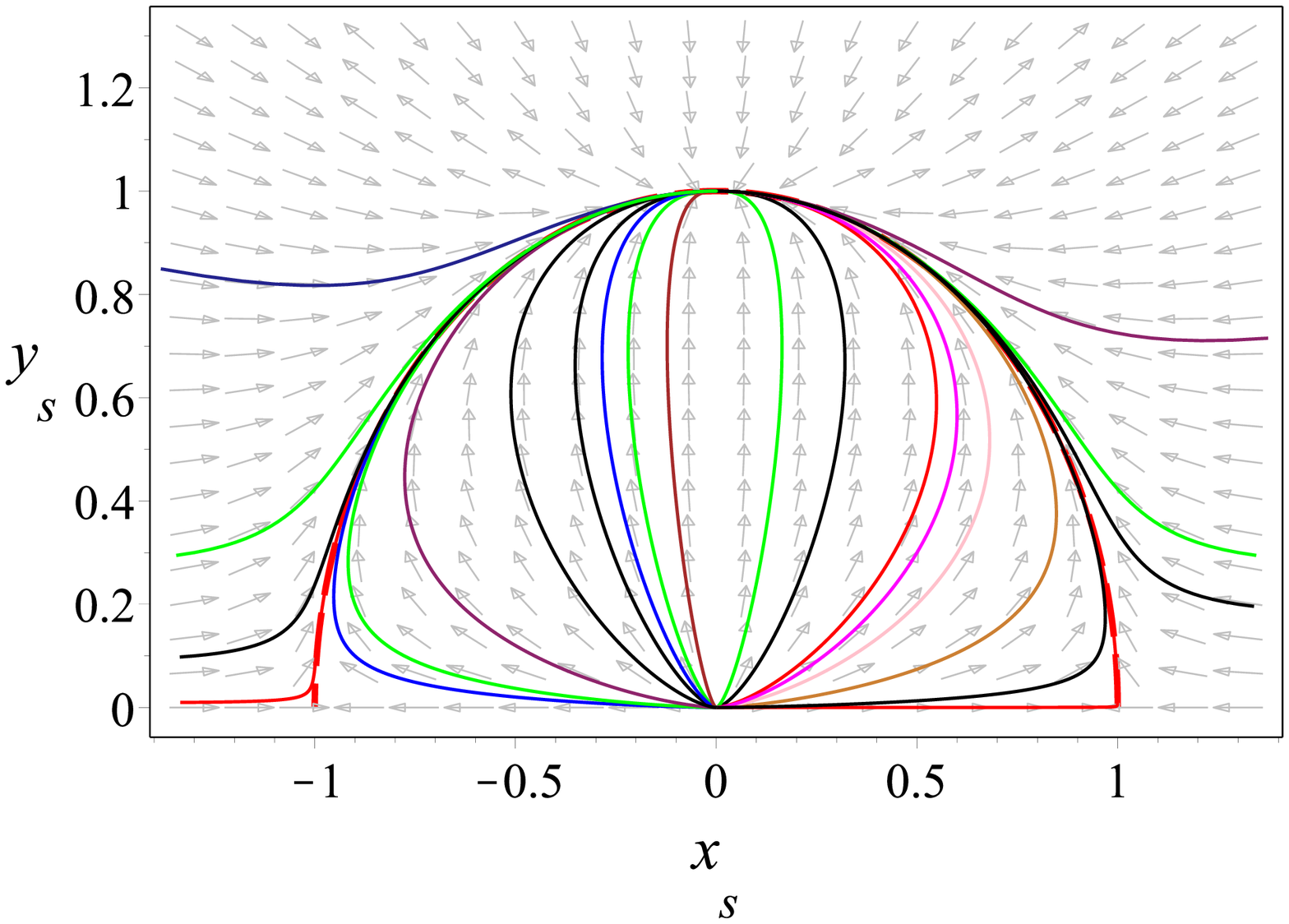}
\includegraphics[width=5cm]{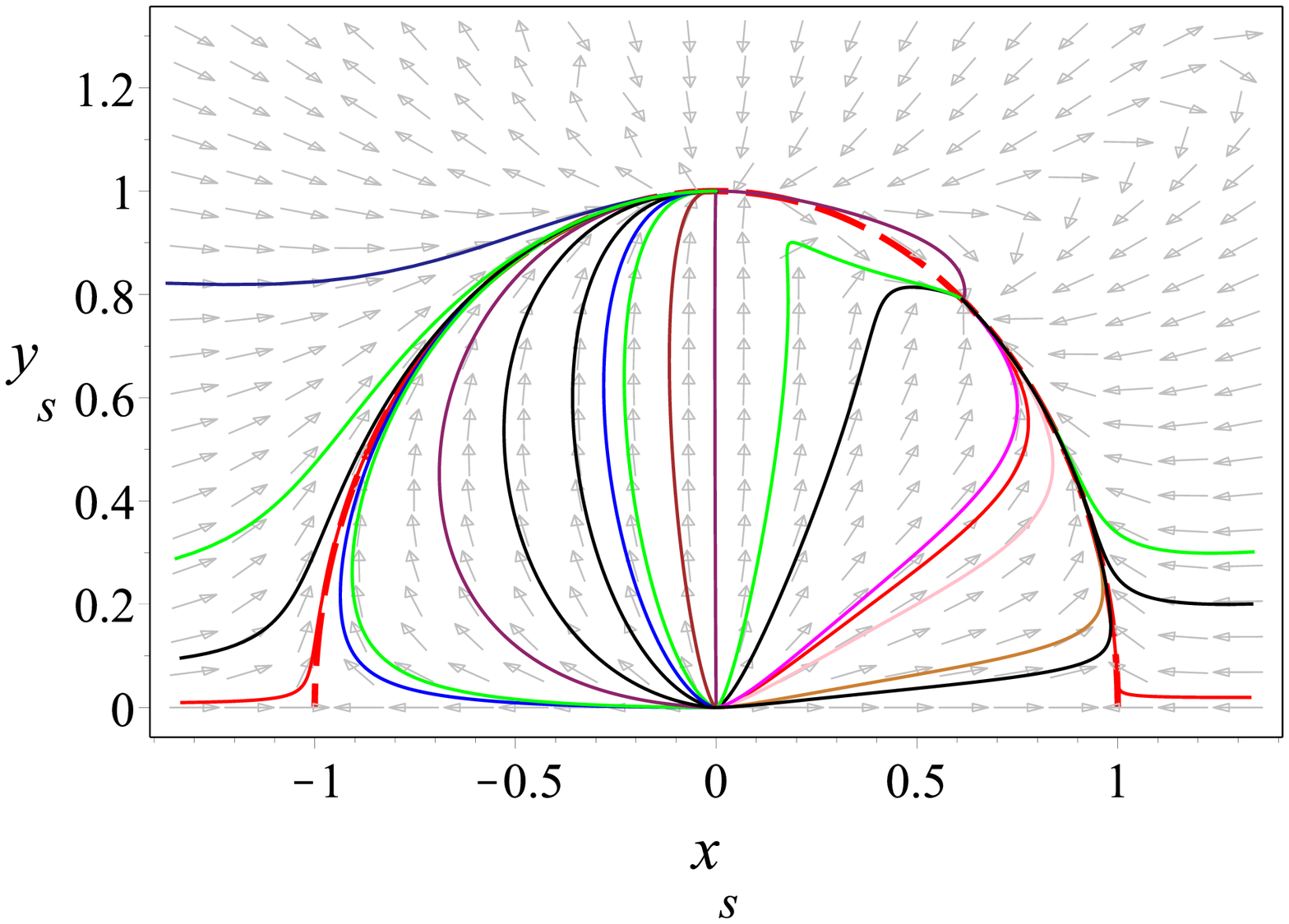}
\includegraphics[width=5cm]{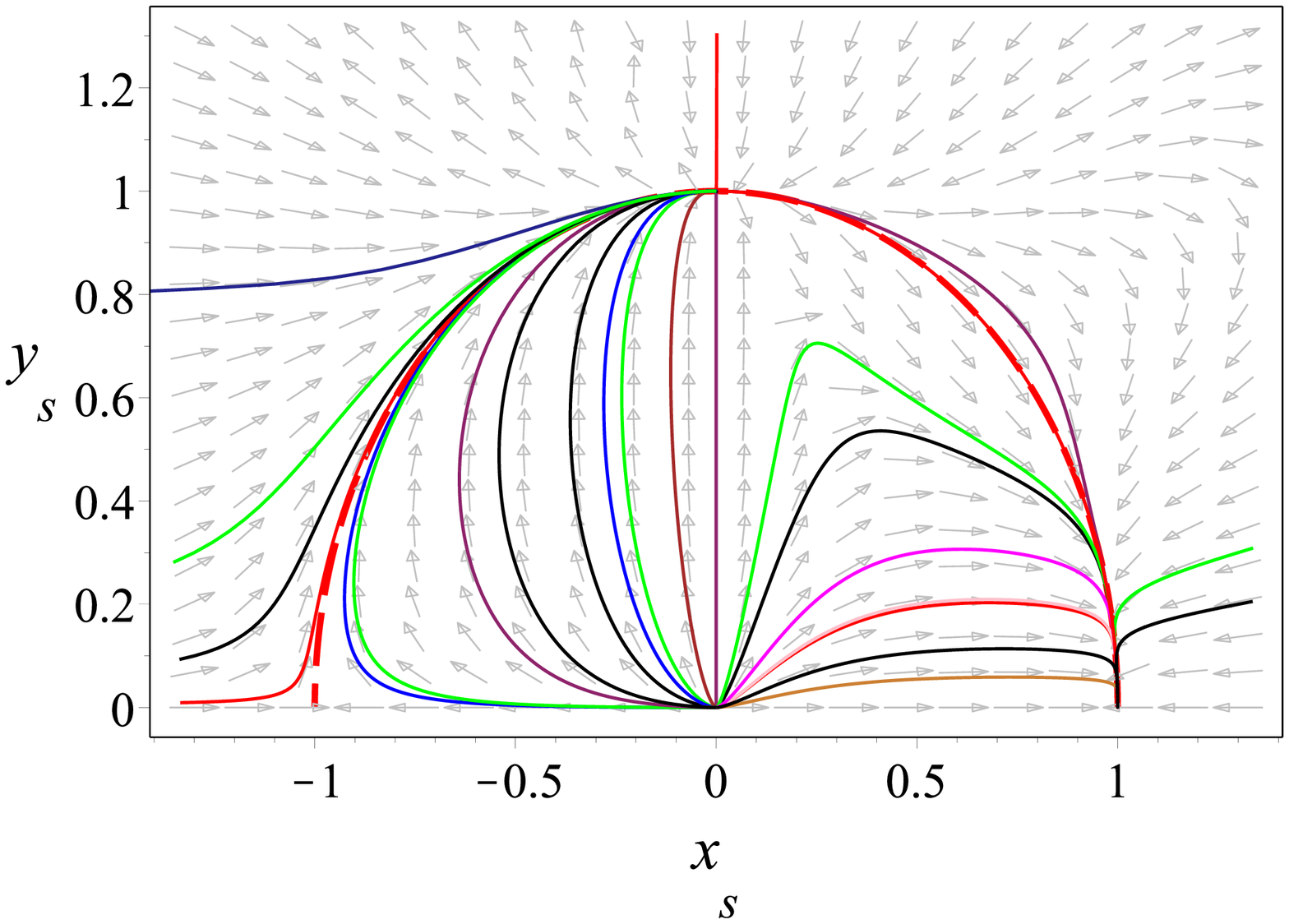}
\includegraphics[width=5cm]{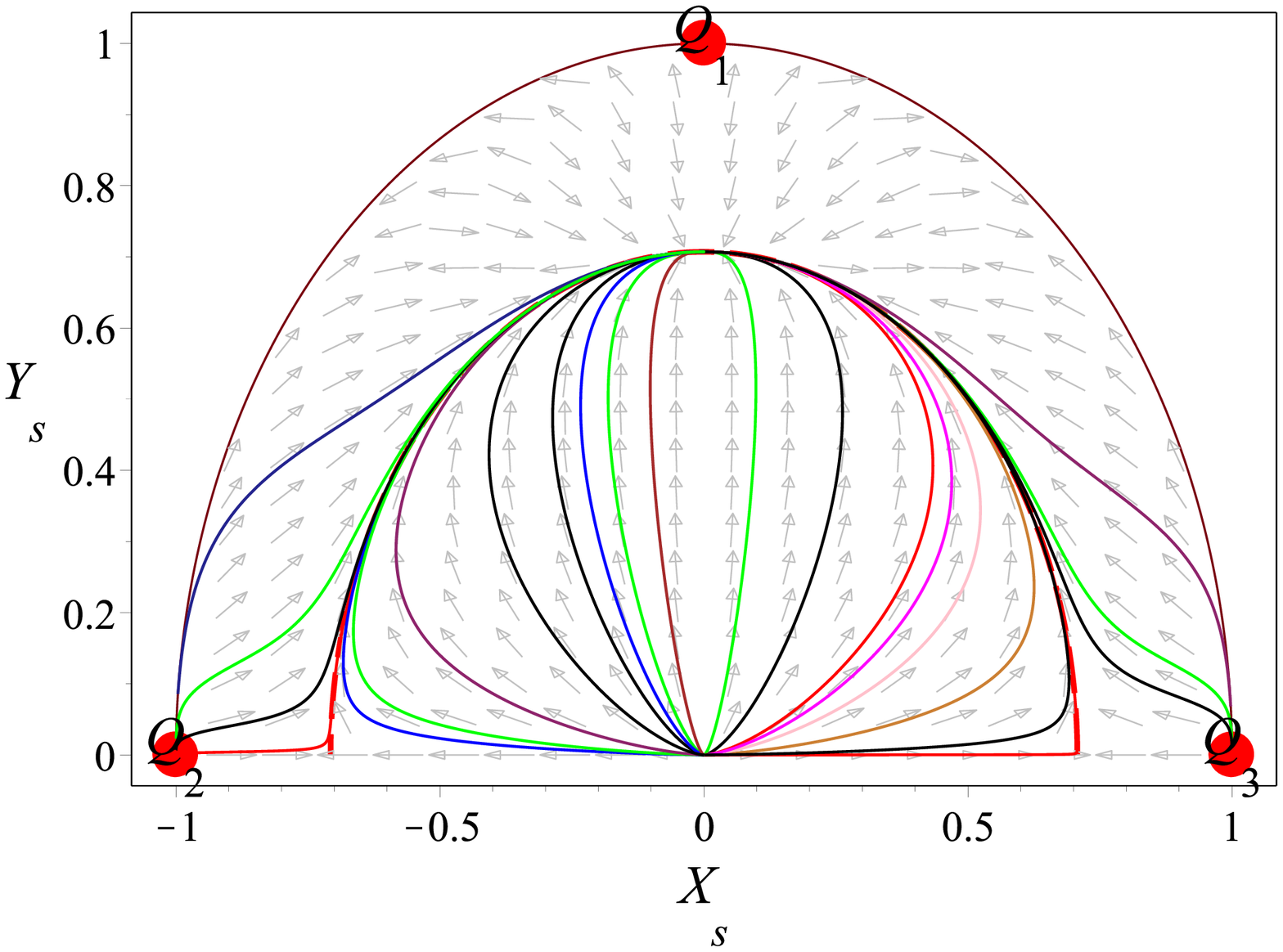}
\includegraphics[width=5cm]{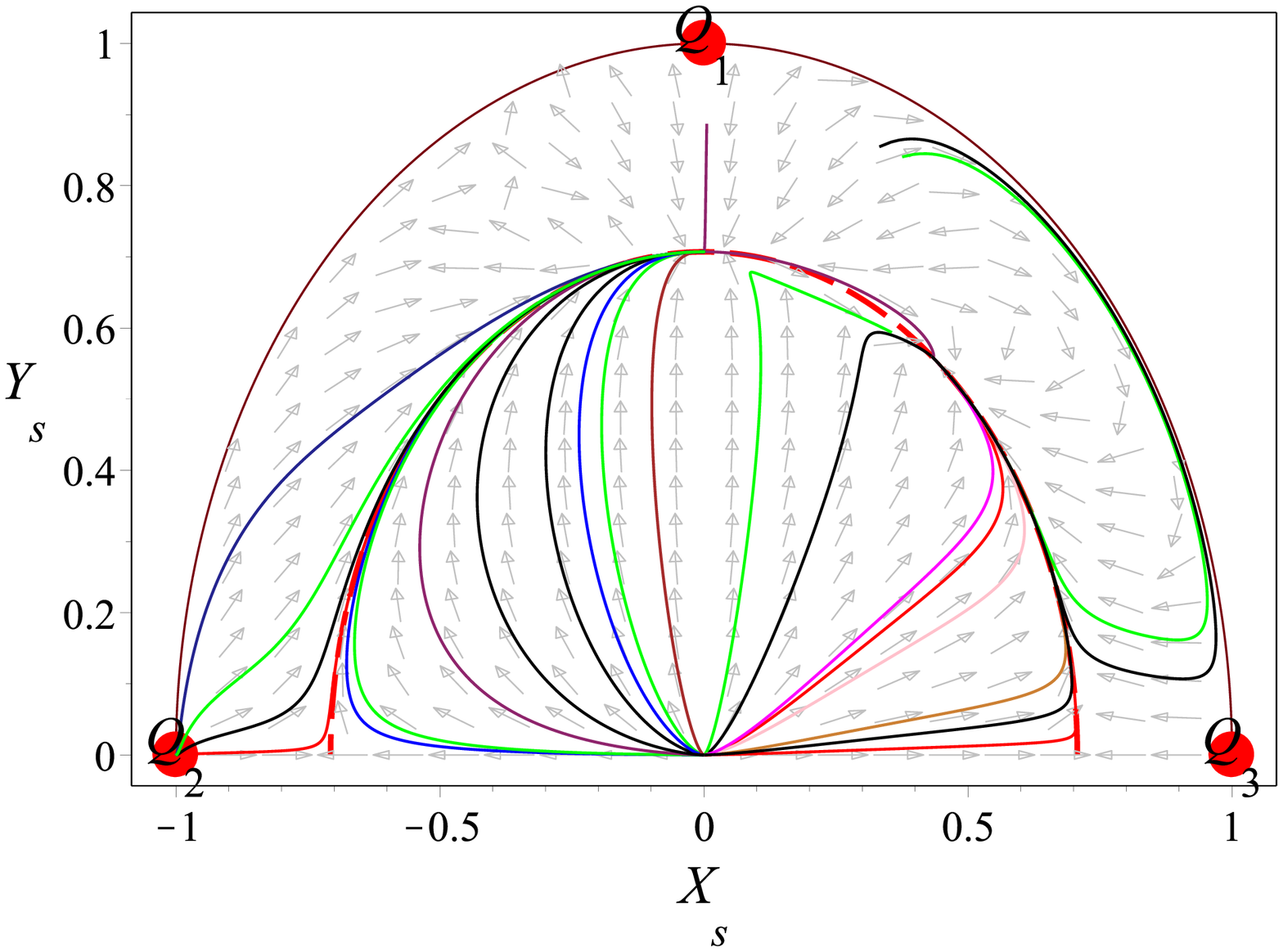}
\includegraphics[width=5cm]{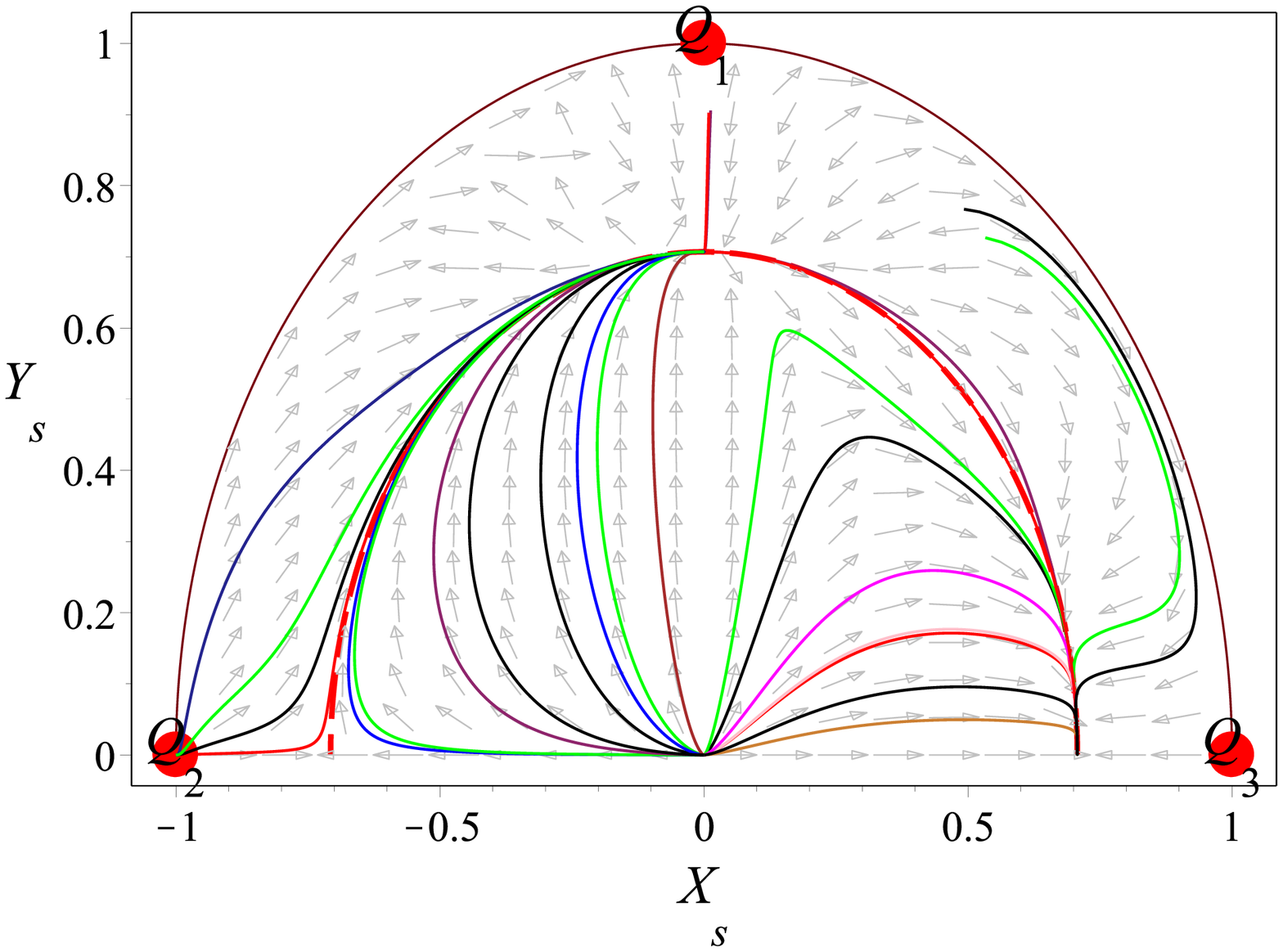}\vspace{0.7cm}
\caption{Phase portrait of the plane-autonomous system of ODE \eqref{asode-g0-exp} for different values of the parameter $\lambda$. From left to right $\lambda=0$ (constant potential), $1.5$ and $3$, respectively. In the bottom panels the compact Poincar\`e phase portrait \eqref{infinity}, which corresponds to the whole (finite and infinite) dynamics, is shown in each case. It is seen that, thanks to the galileon coupling $g_0$, the critical points can be found not only within the semi-disk $x_s^2+y_s^2\leq 1$ (region enclosed by the dashed-curve), but also outside it. There are configurations at infinity corresponding to $x_s\rightarrow 0, y_s\rightarrow +\infty$ ($Q_1$) and to  $x_s\rightarrow \mp \infty, y_s\rightarrow 0$ ($Q_2$ or $Q_3$). }\label{fig-1}
\end{figure*}



\section{Generalized galileon vacuum}\label{sec-vac}

Apparently, the simplest case we can deal with is when the cosmic background is the vacuum ($\Omega_m=0$). In such a case the Friedmann constraint (\ref{friedmann-c}) amounts to a relationship between the variables $x_s$, $y_s$ and $z$:

\bea Q=\frac{1}{2}\sqrt\frac{3}{2}\left(\frac{1-x_s^2-y_s^2}{x_s^3}\right),\label{fr-g0-exp}\eea so that one of these variables, say $z$: $$z=\frac{6\sqrt{6}x_s^3}{6\sqrt{6}x_s^3+x_s^2+y_s^2-1},$$ is redundant, and one ends up with a plane-autonomous system of ODE:

\bea &&x'_s=\frac{1}{\sqrt 6}\frac{\ddot\phi}{H^2}-x_s\frac{\dot H}{H^2},\nonumber\\
&&y'_s=-y_s\left(\sqrt\frac{3}{2}\,\lambda x_s+\frac{\dot H}{H^2}\right),\label{asode-g0-exp}\eea \begin{widetext} where

\bea &&\frac{\dot H}{H^2}=-\frac{6(1-y_s^2)(1+x_s^2-y_s^2)-3(x_s^2+y_s^2-1)(1+x_s^2-y_s^2-\sqrt{2/3}\lambda x_sy_s^2)}{4(1-y_s^2)+(x_s^2+y_s^2-1)^2},\nonumber\\
&&\frac{\ddot\phi}{H^2}=-\frac{3\sqrt{6}\,x_s(x_s^2+y_s^2-1)(1+x_s^2-y_s^2)-6\sqrt{6}\,x_s(1+x_s^2-y_s^2-\sqrt{2/3}\lambda x_sy_s^2)}{4(1-y_s^2)+(x_s^2+y_s^2-1)^2}.\nonumber\eea\end{widetext}

The structure of the dynamical system (\ref{asode-g0-exp}) entails that the semi-infinite planes $\{(x_s,y_s):x_s>0,\,y_s\geq 0\}$ and $\{(x_s,y_s):x_s<0,\,y_s\geq 0\}$ are invariant subspaces. This means, that if one gives initial conditions in one of these subspaces, the corresponding orbits of (\ref{asode-g0-exp}) will entirely lay in that subspace. The vertical lines $x_s=0$, $y_s\geq 0$, and $y_s=0$ are also invariant subspaces. Actually, as seen in the FIG. \ref{fig-1}, the vertical line $x_s=0$ ($y_s\geq 0$) is a separatrix in the phase space. Hence, the orbits originated from initial conditions in the region $\Psi^-=\{(x_s,y_s):x_s<0,\;y_s\geq 0\},$ will lay entirely in this region. The same is true for orbits in the region $\Psi^+=\{(x_s,y_s):x_s>0,\;y_s\geq 0\}.$ Furthermore, the system (\ref{asode-g0-exp}) is form-invariant under the coordinate change $(x_s,y_s,\lambda)\rightarrow (-x_s,y_s,-\lambda).$ Thus, for the numerics we may consider to investigate just the sector $x_s\geq 0, \lambda\geq 0$. The dynamics on the sector $x_s\leq 0,\lambda\leq 0$ will be the same.  
  
Another interesting thing one may read off from FIG. \ref{fig-1} is that, depending on the initial conditions, the phase space orbits may originate either at the infinities $x_s\rightarrow\pm\infty$, or at the big bang $(x_s,y_s)=(0,0)$ $\Rightarrow\;z=0$ (see Eq. (\ref{fr-g0-exp})). This means that the variables $x_s$, $y_s$ are unbounded, which poses a problem for the standard variables (\ref{xy-var}), since one or several critical points at infinity may be lost. 

One possibility is to consider the compact Poincar\`e variables, however, even if these are bounded variables, in the case of interest in this paper, one or several points at infinity may be mapped into a single degenerated point under the Poincar\`e projection (see the demonstration of this fact in the appendix). In spite of this, in the bottom panel of FIG. \ref{fig-1}, the compact Poincar\`e phase portrait of the dynamical system \eqref{asode-g0-exp} is shown for completeness. 

One may naively expect that there can be no new interesting dynamics in the vacuum case with respect to the results of the previous, more general case, where the background matter is considered. In spite of this, in this section we shall investigate the simplest vacuum case and, as we shall show, surprisingly, a new very rich asymptotics arises.

\subsection{New variables}

As mentioned we shall seek for new bounded variables so that all of the possible equilibrium points are ``visible''. Since we renounced to the Poincar\`e variables as long as one or several equilibrium points may be degenerate (see the appendix), here we shall use a different set of variables.

Given that in FIG. \ref{fig-1} the vertical line $x_s=0$, is a separatrix, one may investigate the dynamics in the invariant subspaces $\Psi^-$ and $\Psi^+$, separately. Accordingly, one may introduce the new bounded variables defined in Eq. (\ref{n-var}). The corresponding phase space where to look for equilibrium points: $\Phi_\text{whole}=\Phi^-\cup\Phi^+$, is the union of the following bounded planes: 

\begin{widetext}
\bea \Phi^+=\{(x_+,y):0\leq x_+\leq 1,\;0\leq y\leq 1\},\;\Phi^-=\{(x_-,y):-1\leq x_-\leq 0,\;0\leq y\leq 1\}.\label{phipm}\eea\end{widetext}


\begin{table*}[tbh]\centering
\begin{tabular}{|c||c|c||c|c|c|c|c|c|}
\hline\hline
Crit. Point&\;\;\;$x_\pm$\;\;\;&\;\;\;$y$\;\;\;&Existence&\;\;\;$z$\;\;\;&\;\;\;$q$\;\;\;&\;\;\;$\lambda_1$\;\;\;&\;\;\;$\lambda_2$\;\;\;& Stability \\
\hline\hline
$P^\pm_{1v}$&$\pm 1$ & $0$ & always & $0$ & $8$ & $12$ & $-9$ & saddle \\
\hline
$P^\pm_{2v}$& $\pm 1$ & $1$ & '' & $0$ & $4/5$ & $6/5$ & $9/5$ & unstable \\
\hline
$P^\pm_{3v}$&$\frac{\pm\lambda}{\lambda\mp 2\sqrt{6}}$ & $0$ & $\pm\lambda<0$ & $0$ & $-4$ & $-3$ & $-12$ & stable \\
\hline
$P_{4v}$&$0$ & $1$ & always & $1$ & $-4$ & undef. & $6$ & saddle if $\pm\lambda\geq 0$\\
& & & & & & & & unstable if $\pm\lambda<0$\\
& & & & & & & & (numeric inv.)\\
\hline
$P^\pm_{5v}$&$\pm 1/2$ & $1$& '' & $1$ & $2$ & $3\mp\sqrt\frac{3}{2}\lambda$ & $-6$ & saddle if $\pm\lambda<\sqrt{6}$ \\
 & & & & & & & & stable if $\pm\lambda>\sqrt{6}$ \\
\hline
$P^\pm_{6v}$&$\frac{\sqrt{6}}{\lambda\pm\sqrt{6}}$ & $\frac{\sqrt{6}}{\sqrt{6-\lambda^2}+\sqrt{6}}$& $\lambda^2<6$ & $1$ & $-1+\frac{\lambda^2}{2}$ & $-\lambda^2$ & $-3+\frac{\lambda^2}{2}$ & stable \\
\hline
$P^\pm_{7v}$&$\pm 1$ & $1/2$& always & undet. & $-1$ & $0$ & $-3$ & stable (num. inv.) \\
\hline\hline
\end{tabular}\caption{Critical points of the dynamical system \eqref{asode-nvar}.}\label{tab-2}
\end{table*}

\subsection{The dynamical system}

In terms of the bounded variables defined in (\ref{n-var}), the following plane-autonomous dynamical system is obtained for the galileon vacuum with constant coupling $g=g_0$, and exponential potential $V=V_0\exp(-\lambda\phi)$:

\begin{widetext}
\bea &&x'_\pm=-\frac{x_\pm^2}{\sqrt{6}}\left[\frac{\ddot\phi}{H^2}\right]_\pm+x_\pm(1\mp x_\pm)\left[\frac{\dot H}{H^2}\right]_\pm,\;y'=y(1-y)\left\{\sqrt\frac{3}{2}\lambda\left(\frac{1\mp x_\pm}{x_\pm}\right)+\left[\frac{\dot H}{H^2}\right]_\pm\right\},\label{asode-nvar}\\
&&\left[\frac{\dot H}{H^2}\right]_\pm=-3\left\{\frac{3\Theta_\pm(-1/3)\Theta_\pm(1)-\sqrt{2/3}\lambda x_\pm(1\mp x_\pm)(1-y)^2\Theta_\pm(-1)}{4x^4_\pm y^2(2y-1)+\Theta^2_\pm(-1)}\right\},\nonumber\\
&&\left[\frac{\ddot\phi}{H^2}\right]_\pm=3\left\{\frac{\sqrt{6}(1\mp x_\pm)\Theta_\pm(1)\left[\Theta_\pm(-1)-2x^2_\pm y^2\right]+4\lambda x^3_\pm(1\mp x_\pm)^2y^2(1-y)^2}{x_\pm\left[4x^4_\pm y^2(2y-1)+\Theta^2_\pm(-1)\right]}\right\}.\nonumber\eea\end{widetext} 

Here the '+' and '-' signs refer to two different branches, so that we have in fact two different dynamical systems: i) the one expressed in terms of the variables $x_+$, $y$ and $z$, which corresponds to the case with $\dot\phi>0$, and ii) the other expressed through $x_-$, $y$ and $z$, which corresponds to the case with $\dot\phi<0$. 

In the present case (galileon vacuum), one of the variables: $z$, is expressed as a function of the remaining variables $x_\pm$ and $y$ through 

\bea Q_\pm=9\left(\frac{z-1}{z}\right)=\frac{x_\pm\Theta_\pm(-1)}{2\sqrt{2/3}(1\mp x_\pm)^3y^2},\label{rel-Q}\eea where we have used the definition of the function $\Theta_\pm(a)$ given in Eq. (\ref{defs}).


\begin{figure*}\begin{center}
\includegraphics[width=5cm]{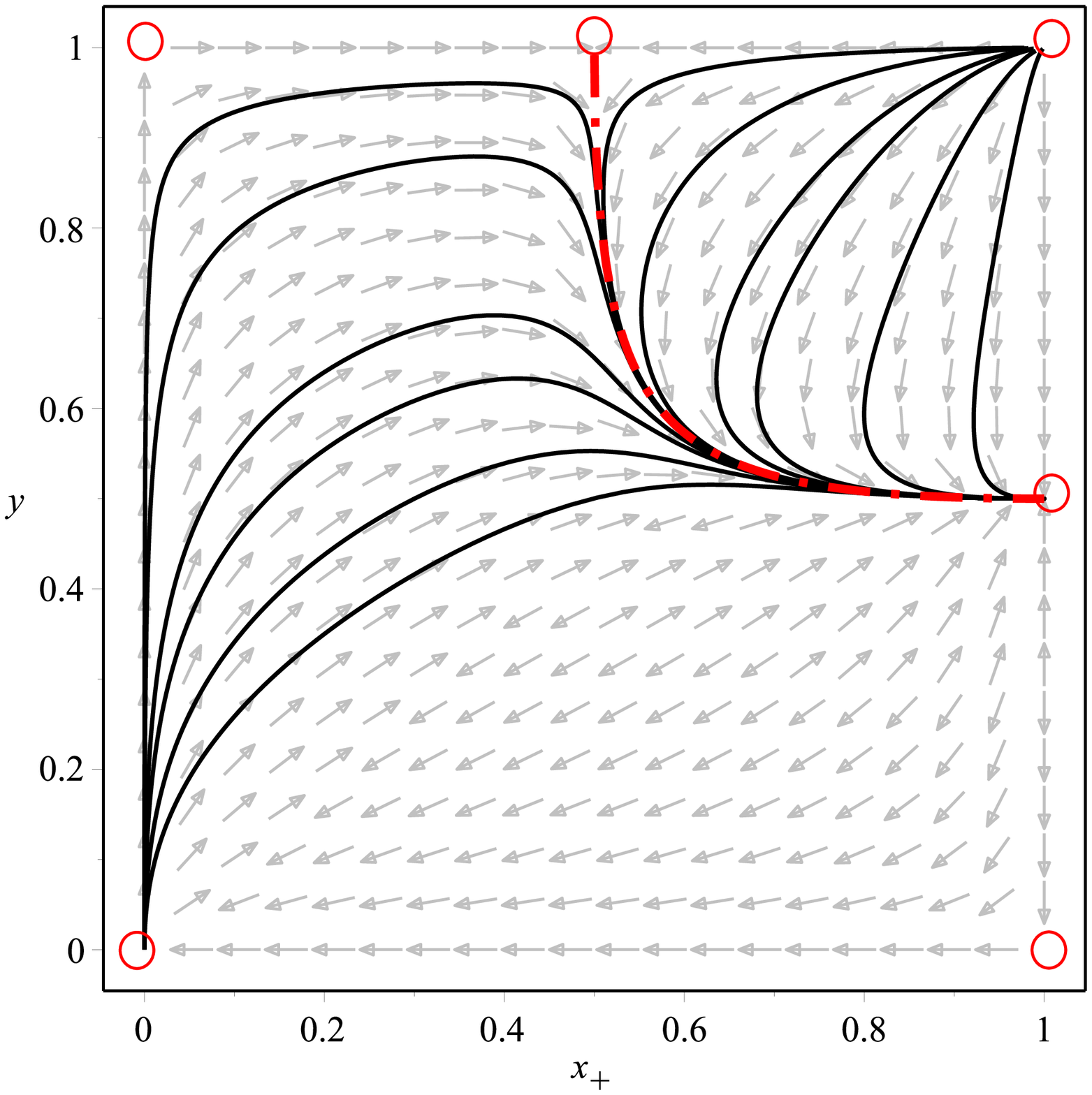}
\includegraphics[width=5cm]{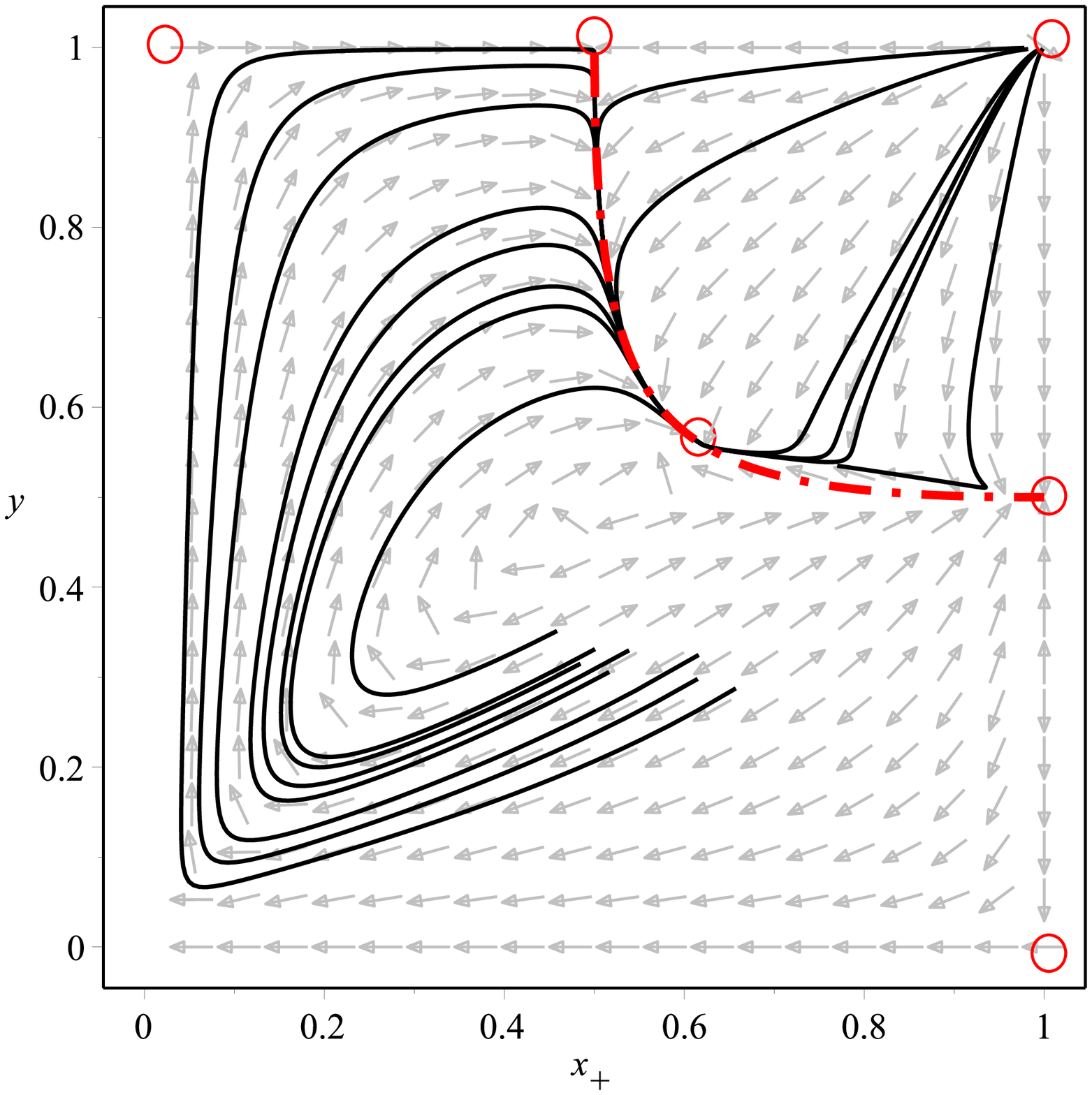}
\includegraphics[width=5cm]{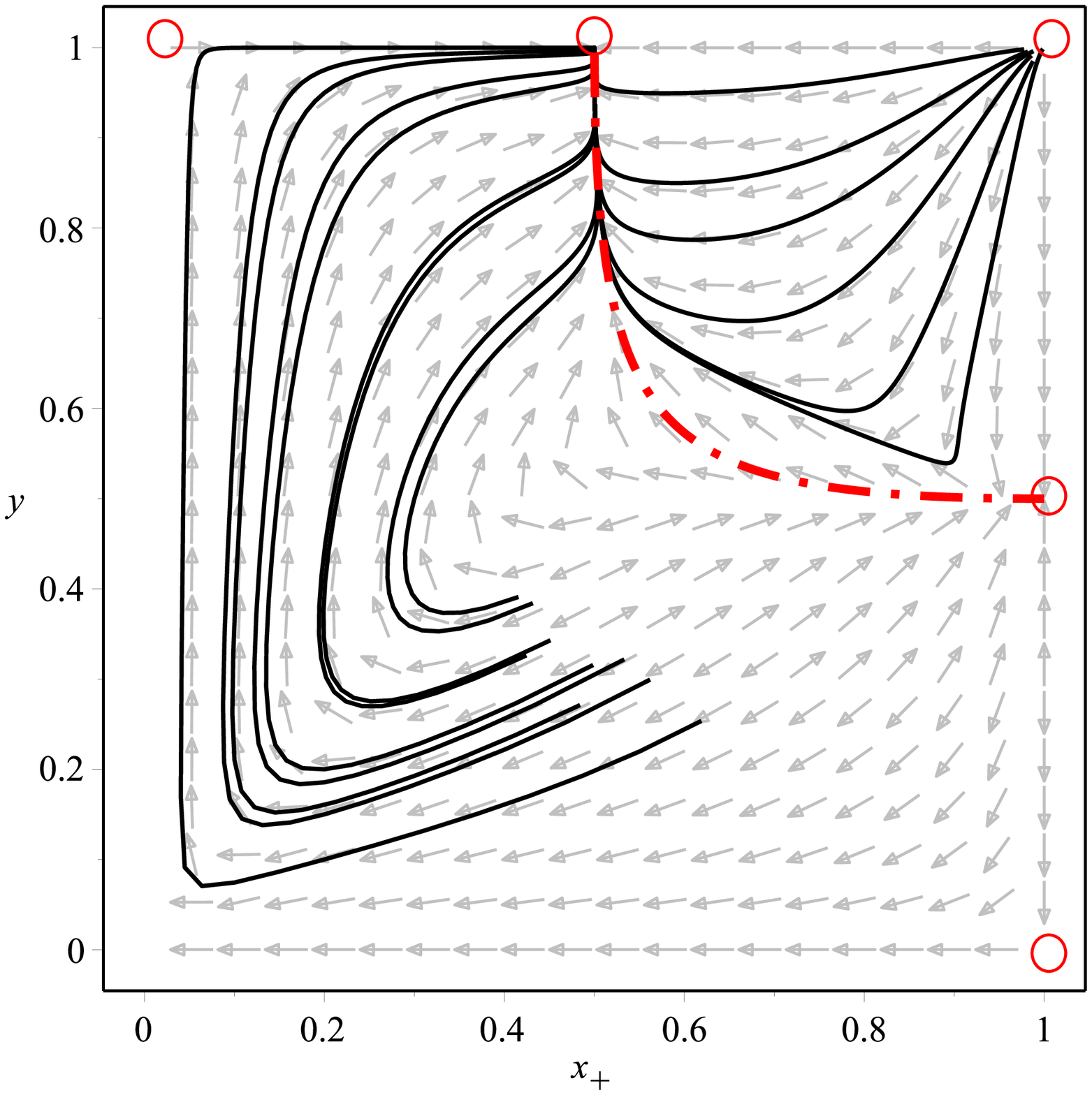}
\includegraphics[width=5cm]{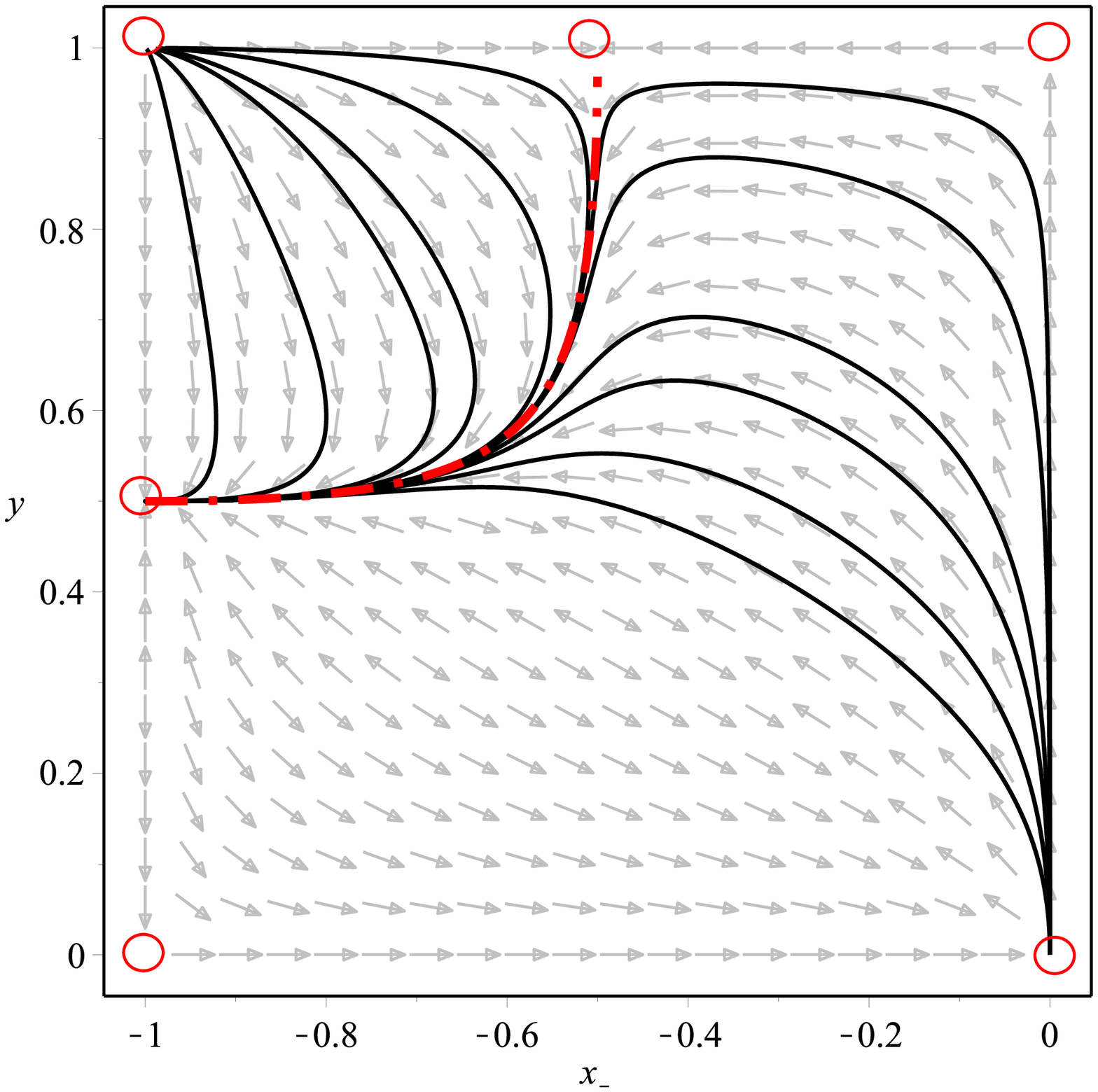}
\includegraphics[width=5cm]{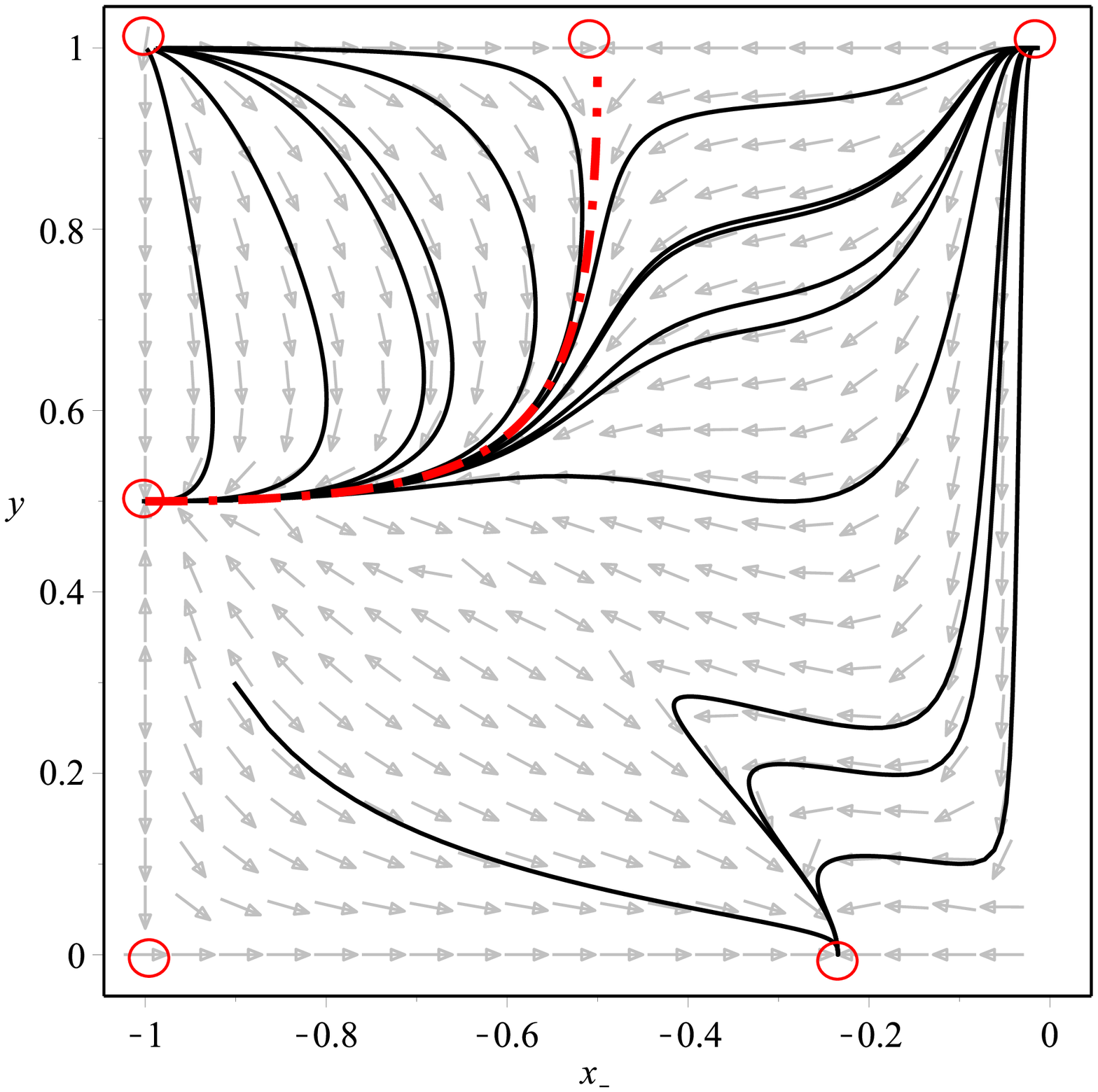}
\includegraphics[width=5cm]{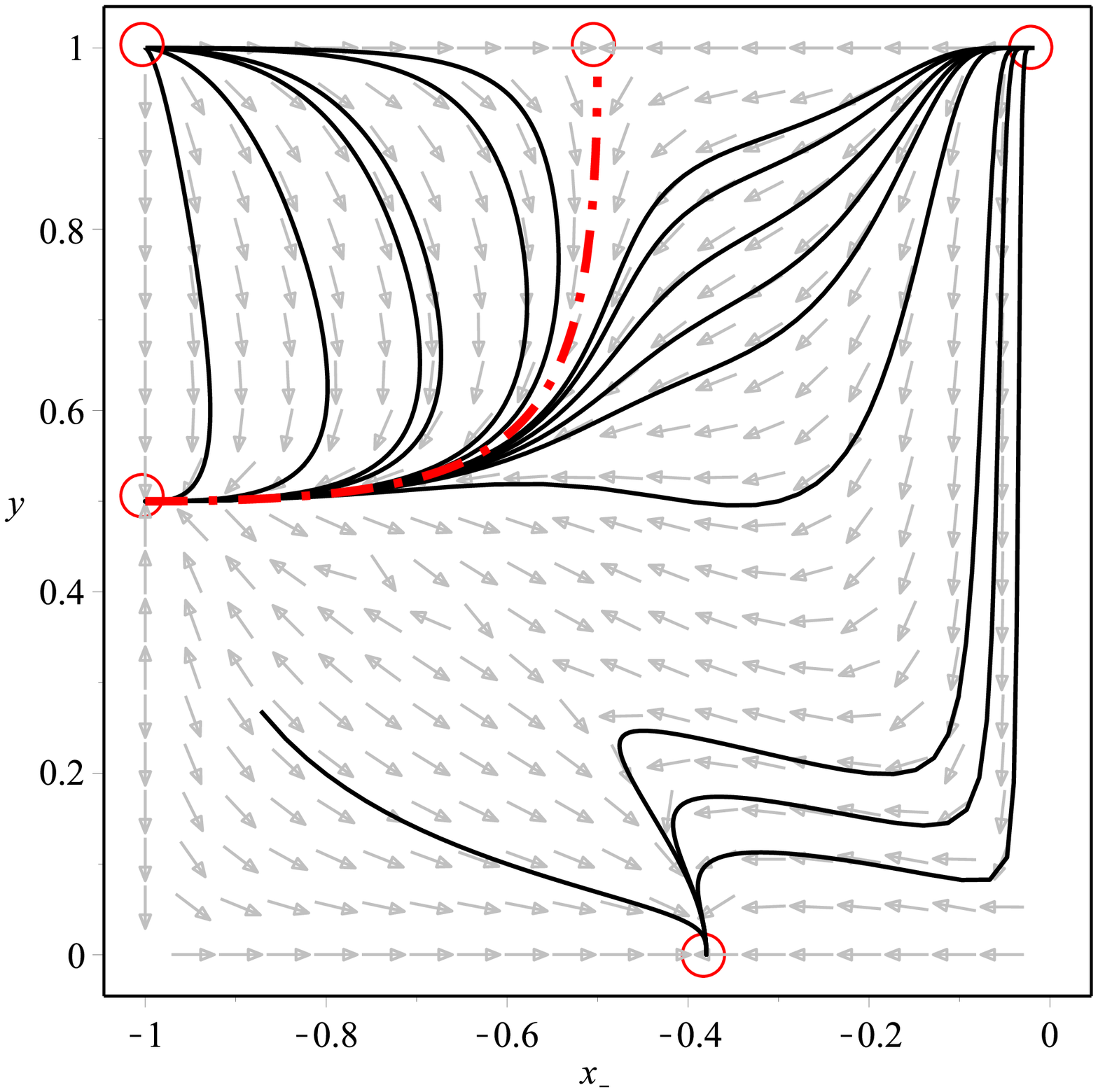}\vspace{0.7cm}
\caption{Phase portrait of the plane-autonomous systems of ODE (\ref{asode-nvar}) for different values of the parameter $\lambda$. From left to right: $\lambda=0$ (constant potential), $1.5$ and $3$, respectively. The top panels are for the phase plane $\{(x_+,y):0\leq x_+\leq 1,\,0\leq y\leq 1\}$, corresponding to the positive branch of \eqref{asode-nvar} ($\dot\phi\geq 0$), while the bottom panels are for the phase plane $\{(x_-,y):-1\leq x_-\leq 0,\,0\leq y\leq 1\}$, which corresponds to the negative branch instead ($\dot\phi\leq 0$). The thick dot-dash curve represents the separatrix $z=1$, which joints the stiff-matter points $(x_+,y)=(1/2,1)$ and $(x_-,y)=(-1/2,1)$, with the de Sitter critical point $(1,1/2)$ and $(-1,1/2)$, respectively. It is seen in the top panels that, depending on the value of $\lambda$, either the de Sitter solution $(1,1/2)$, the DE-dominated solution (critical point on the separatrix sep$^+$), or the stiff-matter solution $(1/2,1)$, can be the late-time attractor. Meanwhile, as seen from the bottom panels, depending on the chosen initial conditions the de Sitter and the phantom solutions can be the late-time attractors.}\label{fig-2}\end{center}\end{figure*}


\subsection{Critical points and phase plane structure}

The whole phase plane for this case is the union of the subspaces $\Phi^+$ and $\Phi^-$ defined in (\ref{phipm}): $\Phi_\text{whole}=\Phi^-\cup\Phi^+$. Its boundaries are at the edges 

\begin{align} 
&B_1:=\{(x_-,0): -1\leq x_-\leq 0\}\cup \{(x_+,0): 0\leq x_+\leq 1\},\nonumber \\ 
& B_2:=\left\{(1,y): -\infty \leq  y \leq \infty\right\},\nonumber\\
&B_3:=\{(x_-,1): -1\leq x_-\leq 0\}\cup \{(x_+,1): 0\leq x_+\leq 1\},\nonumber \\
&B_4:=\left\{(-1,y): -\infty \leq  y \leq \infty\right\},\label{boundaries}
\end{align} where $z=0$ $\Leftrightarrow$ $g_0H^2\rightarrow\infty$, which means that either there is a cosmological singularity there ($H\rightarrow\infty$), or the cubic derivative interaction is decoupled from the gravitational interactions ($g_0\rightarrow\infty$). 

The following separatrices can be identified:

\bea &&\text{sep}^0:=\left(0,y\right),\nonumber \\
&&\text{sep}^+:=\left(x_+,\frac{x_+}{x_++\sqrt{2x_+-1}}\right),\nonumber \\
&&\text{sep}^-:=\left(x_-,\frac{x_-}{x_--\sqrt{-2x_--1}}\right).\label{separatrix}\eea Notice that on the separatrices sep$^0$ and sep$^\pm$, $z=1$, i. e., $g_0 H^2=0$, so that either we deal with the static universe there ($H=0$), or, if $g_0=0$, the standard quintessence model with exponential potential (basically Einstein's general relativity plus a fluid of self-interacting scalar field with exponential potential \cite{wands}), is recovered.

\subsubsection{Exponential quintessence}

The critical points of the dynamical systems (\ref{asode-nvar}), together with their main properties, are shown in TAB. \ref{tab-2}. The equilibrium points $P_{5v}^\pm$ and $P^\pm_{6v}$, for which $z=1$, are the usual critical points found in Ref. \cite{wands} for the exponential quintessence model, if consider, as we do in the present section, the vacuum case (no other matter degrees of freedom than the scalar field). 

The points $P_{5v}^\pm$ correspond to the stiff-matter solutions, which, in the exponential quintessence case, are unstable and are expected to be relevant only at early times \cite{wands}. In the present case, however, we obtain a bit different result which is due to the non-vanishing galileon coupling $g=g_0$. Actually, as seen from TAB. \ref{tab-2} and also from the phase portrait in the FIG. \ref{fig-2} (top right-hand panel), the stiff-matter solution can be stable, i. e., it can be a late-time attractor. This is achieved if either, $\lambda>\sqrt{6}$ ('+' branch), or $\lambda<-\sqrt{6}$ ('-' branch). For either $\lambda<\sqrt{6}$ ('+' branch) or $\lambda>-\sqrt{6}$ ('-' branch), the stiff-matter solution is a saddle point in the phase space (see FIG. \ref{fig-2}), but it can not be a source point (past attractor) as it is in the standard exponential quintessence case if $|\lambda|<\sqrt{6}$. This apparently harmless departure from the standard stability properties of the stiff-matter solution arises because the equilibrium point: $x_\pm=\pm 1/2$, $y=1$ ($z=1$), or, in terms of the standard variables $x_s$, $y_s$ in Eq. (\ref{xy-var}) (the same variables used in \cite{wands}): $x_s=\pm 1$, $y_s=0$, is approached asymptotically not only if $\dot\phi=\pm\sqrt{6}H$, $V=0$, $g_0=0$, as in the quintessence case, but also if there is a perhaps very tiny residual non-vanishing galileon coupling $g_0\neq 0$ ($g_0\ll 1$): $$\dot\phi\sim H\gg V,\;g_0\ll 1/H^2,\;g_0\neq 0.$$ Hence, provided that $|\lambda|>\sqrt{6}$, and that the above conditions are fulfilled, the stiff-matter solution is the global attractor (see the top right-hand panel of FIG. \ref{fig-2}), meaning that the final (stable) state of the cosmic evolution is the ultra-relativistic stiff-matter stage. This behavior has not analogue in the exponential quintessence model.

The critical points $P^\pm_{6v}$ have the same properties as in the exponential quintessence model \cite{wands}. These correspond to scaling of the kinetic and potential energies of the scalar field: $$\frac{\dot\phi^2}{2V}=\frac{\lambda^2}{6-\lambda^2}.$$ Whenever they exist, they are attractors.

\subsubsection{The de Sitter solution}\label{subsec-no-self-a}

Another interesting property of the present galileon model, formerly investigated in \cite{genly}, is that the de Sitter solution (points $P_{7v}^\pm$ in TAB. \ref{tab-2}) is a critical point of (\ref{asode-nvar}). It is a local attractor.\footnote{Worth noticing that the points $P_{5v}^\pm$ and $P_{6v}$, correspond to the points $A^\pm$ and $C$ in table 1 of \cite{genly}, respectively.} The parameter $z$ is undefined in this case since, if the de Sitter point is approached along the separatrices sep$^\pm$, then $z=1$, meanwhile, for other approaching directions $z=0$. The de Sitter solution does not arise in standard exponential quintessence, unless $\lambda=0$ (constant potential case), so that its existence for any $\lambda\neq 0$ is a genuine consequence of the galileon coupling $g_0\neq 0$.

As long as the former critical point exists independent on the value of the parameter $\lambda$, one might incorrectly infer that for vanishing potential, i. e., in the limit $\lambda\rightarrow\infty$, the equilibrium point $P_{7v}^\pm$ could be associated with a self-accelerating solution as in \cite{silva_koyama}, i. e, a de Sitter solution in which cosmic acceleration arises even in the absence of matter and for vanishing potential: $$\rho_m=p_m=V(\phi)=\dot H=0.$$ 

In \cite{silva_koyama} this kind of solution has been investigated within the context of BD theory with the cubic derivative interaction $\propto f(\phi)(\der\phi)^2\nabla^2\phi$, so that it were not that surprising if this critical point arose in the present case, where a similar cubic interaction is being considered. However, as we shall show, the self-accelerating solution can not arise in the present case. Actually, supposing that the conditions for a self-accelerating solution are fulfilled, i. e., assuming that $\rho_m=p_m=0$, and $\dot H=0$ $\Rightarrow$ $H=H_0$, the Friedmann equation in \eqref{feqs}, amounts to the following cubic algebraic equation in $\dot\phi$: $$9g_0H_0\dot\phi^3-\dot\phi^2+9H_0^2=0.$$ Any real root $\dot\phi=r_0=$const. of this equation leads to $\ddot\phi=0$, hence, the Raychaudhuri equation $$-2\dot H=\rho_\phi+p_\phi=0\;\Rightarrow\;1-3g_0H_0r_0=0.$$ Exactly the same result: $3g_0H_0r_0=1$, is obtained from the Klein-Gordon equation in \eqref{kg-eq}. Now, if substitute this $r_0$ back into the cubic algebraic equation above, one gets that $H^4_0=-2/27g_0^2$, which can not be satisfied by any reals $H_0$ and $g_0$.

\subsubsection{The big bang solution}

The points $P_{2v}^\pm:(\pm 1,1)$ should not be confounded neither with the point $P^\pm_2$ in TAB. \ref{tab-1} ($O_1$ in Ref. \cite{genly}), nor with $P^\pm_1$ in that table. In terms of the standard variables $x_s$, $y_s$, $P_{2v}^\pm\Rightarrow (0,0)$, which, in the case of the exponential quintessence model explored in Ref. \cite{wands}, coincides with $P^\pm_2$ and represents the matter-dominated solution. However, in the vacuum case, since $$\Omega_\phi=\frac{\rho_\phi}{3H^2}=1,$$ at any time, it can not represent any matter-dominance. In fact, since for $P_{2v}^\pm$, $q=4/5$, then $$\frac{\dot H}{H^2}=-\frac{9}{5}\;\Rightarrow\;H=\frac{5/9}{t-t_0}\;\Rightarrow\;a(t)\propto(t-t_0)^{5/9},$$ i. e., $P_{2v}^\pm$ is associated with a solution with a pure galileon big bang singularity at some initial time $t_0$ (compare with the point $P^\pm_1$ in TAB. \ref{tab-1} which is associated with a matter-dominated big bang instead). This unstable solution, which corresponds to a source critical point in the phase space, has not analogues in the standard exponential quintessence model. It can have importance only at early stages of the cosmic evolution.

\subsubsection{The phantom solution}

One of the most interesting findings of the present investigation is the solution which is associated with the critical point $P_{3v}^\pm$. It is a stable solution (a local attractor) and represents phantom behavior. In order to illustrate the latter statement let us to choose the $P_{3v}^+$ solution, which exists only for negative $\lambda<0$. Let us set $\lambda=-\kappa$, with $\kappa>0$. At $P_{3v}^+$ we have that 

\bea &&x_+=\frac{\kappa}{\kappa+2\sqrt{6}}\;\Rightarrow\;\dot\phi=\frac{12}{\kappa}\,H,\nonumber\\
&&y=0\;\Rightarrow\;\frac{\sqrt{V}}{\sqrt{3}H}\rightarrow\infty,\nonumber\\
&&z=0\;\Rightarrow\;g_0H^2\rightarrow\infty.\nonumber\eea 

From the first equation above it follows that $$\phi(a)=\frac{12}{\kappa}\,\ln a+\phi_0,$$ where $\phi_0$ is an arbitrary integration constant. Additionally, since for this critical point $q=-4$ (recall that for the vacuum case $\Omega_\phi=1$): $$\frac{\dot H}{H^2}=-\frac{3}{2}\left(1+\omega_\phi\right)=3\;\Rightarrow\;\omega_\phi=-3,$$ where, by definition $\omega_\phi:=p_\phi/\rho_\phi$, the associated solution is a super-accelerating one. For this case we have that $$H(t)=\frac{1}{3(t_f-t)}\;\Rightarrow\;a(t)=\frac{a_0}{(t_f-t)^{1/3}}\;(t\leq t_f),$$ where $-3t_f$ and $\ln a_0$ are arbitrary integration constants. 

Given that $V\propto\exp(\kappa\phi)\propto a^{12}\propto(t_f-t)^{-4}$, then, as $t\rightarrow t_f$ asymptotically: $$H^2g_0\propto(t_f-t)^{-2}\rightarrow\infty,\;\frac{\sqrt V}{H}\propto(t_f-t)^{-1}\rightarrow\infty,$$ as required. We have, also, that $$\rho_\phi(t)=3H^2(t)=\dot H(t)=\frac{1}{3(t_f-t)^2}.$$ Besides, since at $P^\pm_{3v}$, $\dot\phi=12H/\kappa$, the Friedmann equation can be written as $$V=\left(3-\frac{\alpha^2}{2}\right)H^2+3\alpha^3 g_0 H^4,$$ where $\alpha=12/\kappa$. As seen, the self-interaction galileon potential $V(\phi)$ asymptotically approaches to $V\propto H^4$, as required by the consistency of the phantom solution. 

The phantom behavior is evident from the fact that the energy density of the galileon unboundedly grows up with $t$. As seen, given that $a(t)$, $H(t)$, $\dot H(t)$, and $\rho_\phi(t)$, all blow up at $t=t_f$, i. e., in a finite time into the future, a big rip singularity \cite{ruth} is the inevitable fate of the cosmic evolution in the present case.

The point $P_{4v}$ represents super-accelerating contraction of the universe as we shall see below. In contrast to $P_{3v}^\pm$, this solution has no impact in the late-time dynamics. In this case it is required a vanishing self-interaction potential $V=0$ $\Rightarrow\;y=1$, and a finite $\dot\phi\neq 0$, and that, asymptotically, $$H\rightarrow 0\;\Rightarrow\;x_\pm\rightarrow 0,\;z\rightarrow 1.$$ As a matter of fact $$q=-4\;\Rightarrow\;\frac{\dot H}{H^2}=3\;\Rightarrow\;H(t)=-\frac{1}{3(t-t_b)}\;(t\geq t_b),$$ where the integration constant has been set $C=-3t_b$. Asymptotically, as $t\rightarrow\infty$, $H\rightarrow 0$, as required, besides $a(t)\propto(t-t_b)^{-1/3}\rightarrow 0$. Although we restricted ourselves to consider expanding cosmologies only, this point belongs in the boundary of the phase space and so, in spite of the mentioned restriction, we have taken it into consideration in our analysis.

There is yet another pair of equilibrium points of the dynamical system corresponding to the generalized galileon vacuum, which are not found in the more general case when, in addition to the galileon field, there is standard (pressureless) matter in the cosmic background. These are the points $P^\pm_{1v}$ in TAB. \ref{tab-2}. They correspond to a super-decelerated pace of the cosmic expansion $$H\propto\frac{1}{9(t-t_0)}\;\Rightarrow\;a(t)\propto(t-t_0)^{1/9},$$ where $9t_0$ is an arbitrary integration constant. At these points: $$\dot\phi\ll H\ll\sqrt{V},\;1\ll \sqrt{g_0} H\ll(\sqrt{g_0}\dot\phi)^3.$$ The points $P^\pm_{1v}$ are saddle critical points, so that the corresponding pattern of cosmic expansion can be only a transient stage of the cosmological evolution. Besides, as seen from FIG. \ref{fig-2}, only for e very narrow set of initial conditions the corresponding phase plane orbits approach to $P^\pm_{1v}$.

As shown, the asymptotic structure of the present galileon model, which is a particular case of the model with matter, is not as trivial as thought. In particular the galileons can play an important role in determining the fate of the cosmic evolution. This is to be contrasted with the result of section \ref{sec-mat} (see Ref. \cite{genly}) that, in the presence of background matter, the galileons will not have impact on the late-time evolution of the universe.


\begin{table*}[tbh]\centering
\begin{tabular}{|c||c|c||c|c|c|}
\hline\hline
Crit. Point&\;\;\;$x_s$\;\;\;&\;\;\;$y_s$\;\;\;&Existence&\;\;\;Stability\;\;\;&\;\;\;$\Omega_m$\;\;\;\\
\hline\hline
$P_\text{mat}$&$0$ & $0$ & always & saddle & $1$ \\
\hline
$P^\pm_\text{stiff}$& $\pm 1$ & $0$ & '' & unstable if $\pm\lambda<\sqrt{6}$ & $0$ \\
& & & & saddle if $\pm\lambda>\sqrt{6}$ & \\
\hline
$P_\phi$& $\lambda/\sqrt{6}$ & $\sqrt{1-\lambda^2/6}$ & $\lambda^2<6$ & stable if $\lambda^2<3$ & $0$ \\
& & & & saddle if $3<\lambda^2<6$ & \\
\hline
$P_\text{scaling}$& $\sqrt{3/2}/\lambda$ & $\sqrt{3/2}/\lambda$ & $\lambda^2>3$ & stable node if $3<\lambda^2<24/7$ & $1-3/\lambda^2$ \\
& & & & stable spiral if $\lambda^2>24/7$ & \\
\hline\hline
\end{tabular}\caption{Critical points of the dynamical system (\ref{wands-asode}), corresponding to exponential quintessence \cite{wands}.}\label{tab-3}
\end{table*}


\section{discussion}\label{sec-discuss}

In order to understand the problem we are facing, let us to briefly expose the results of the dynamical systems study of the exponential quintessence model (see Ref. \cite{wands} for details). In this case the standard variables $x_s$, $y_s$ defined in Eq. (\ref{xy-var}) are already bounded phase space variables. Consequently, all of the existing equilibrium points will be located in a bounded region of the phase plane. 

Considering, just for simplicity, pressureless matter ($p_m=0$), the autonomous dynamical system corresponding to this case reads:

\bea &&x'_s=-3x_s+\frac{3}{2}\,x_s\left(1+x_s^2-y_s^2\right)+\sqrt\frac{3}{2}\,\lambda y_s^2,\nonumber\\
&&y'_s=-\sqrt\frac{3}{2}\,\lambda x_sy_s+\frac{3}{2}\,y_s\left(1+x_s^2-y_s^2\right),\label{wands-asode}\eea while the Friedmann constraint is written as:

\bea \Omega_m=1-x_s^2-y_s^2.\label{wands-friedmann-const}\eea 

The critical points $P^*:(x^*_s,y^*_s)$ of the dynamical system (\ref{wands-asode}), which are located within the upper semi-disk $x^2_s+y^2_s\leq 1$ ($y_s\geq 0$), are shown in TAB. \ref{tab-3}. There are two equilibrium points that are associated with the presence of standard mater (other than the scalar field $\phi$): i) the matter dominated solution $P_\text{mat}$ where $\Omega_m=1$, and ii) the scalar field-matter scaling solution $P_\text{scaling}$, with $\Omega_m=1-3/\lambda^2$. Since $\Omega_m=0$, for the remaining critical points: iii) the stiff matter solutions $P^\pm_\text{stiff}$, and iv) the scalar field dominated solution $P_\phi$, these arise even in the absence of standard matter. 

Hence, what one expects for the vacuum of the above theory is that the asymptotic structure in the phase plane will be characterized by the equilibrium points $P^\pm_\text{stiff}$ and $P_\phi$ exclusively. This is corroborated by setting $\Omega_m=0$ in Eq. (\ref{wands-friedmann-const}): $$\Omega_m=0\;\Rightarrow\;x_s^2+y_s^2=1.$$ This relationship between the variables $x_s$ and $y_s$ leads to a reduction of the dimensionality of the dynamical system from 2 to 1. In other words, one is left with a single autonomous ODE: $$x'_s=\left(\sqrt\frac{3}{2}\,\lambda-3x_s\right)\left(1-x_s^2\right).$$ 

Just as expected: the only critical points of this autonomous ODE are $x_s=\pm 1$ ($y_s=0$), and $x_s=\lambda/\sqrt{6}$ ($y_s=\sqrt{1-\lambda^2/6}$), respectively. In consequence, only the critical points $P^\pm_\text{stiff}$ and $P_\phi$, survive in the particular vacuum case.

In the galileon model depicted by the cosmological equations (\ref{feqs}), the situation is a bit more complex. In this case one has to migrate to another set of bounded variables \eqref{n-var}: $$x_\pm=\frac{1}{x_s\pm 1},\;y=\frac{1}{y_s+1},$$ since the standard variables $x_s$ and $y_s$ are unbounded in this model. 

The critical points $P^\pm_*:(x_\pm^*,y^*,z^*)$ of the 3D autonomous system corresponding to the mixture of the galileon and a matter fluid depicted by Eq. \eqref{mat-asode}, are shown in TAB. \ref{tab-1}. Of them, only for $P^\pm_3:\left(\pm 1/2,1,1\right)$ -- the stiff-matter solution, and for $$P^\pm_4:\left(\frac{\sqrt{6}}{\lambda\pm\sqrt{6}},\frac{\sqrt{6}}{\sqrt{6-\lambda^2}+\sqrt{6}},1\right),$$ which is associated with the dark energy-dominated solution, the normalized (dimensionless) energy density of standard matter vanishes: $\Omega_m=0$. Hence, one naively expects that in the galileon vacuum case, i. e., in the particular case when $\Omega_m=0$, only these points will remain. 

The very interesting thing here is that, in addition to $P^\pm_3$ and $P^\pm_4$, which correspond to the critical points $P^\pm_{5v}$ and $P^\pm_{6v}$ in TAB. \ref{tab-2}, respectively, a variety of new equilibrium points $P^\pm_*:(x^*_\pm,y^*)$, is found in the phase plane of the dynamical system (\ref{asode-nvar}), which do not arise when, in addition to the galileon field, standard matter fields populate the cosmic background. Among them the most interesting ones are:

\begin{enumerate}

\item The big bang solution $P^\pm_{2v}:(\pm 1,1)$, is the source point for every orbit to the right and above of the separatrix sep$^+$ ('+' branch of the dynamical system (\ref{asode-nvar})), and to the left and above of the separatrix sep$^-$ ('-' branch of the dynamical system). This is illustrated in the figure \ref{fig-2}.

\item The de Sitter solution $P^\pm_{7v}:(\pm 1,1/2)$, is a local late-time attractor, where for local we mean that there may coexist other late-time attractors.

\item The super-accelerated phantom solution $$P^\pm_{3v}:\left(\frac{\pm\lambda}{\lambda\mp 2\sqrt{6}},0\right),$$ exists only for negative $\lambda<0$, in the '+' branch of (\ref{asode-nvar}), or for positive $\lambda>0$, in the '-' branch. This is also a local late-time attractor (see the bottom center and right-hand panels of FIG. \ref{fig-2}).

\end{enumerate} 

All of the above points which are not found in the more general situation when $\Omega_m\neq 0$, are either late-time or past attractors, so these can have impact in the asymptotic future and/or past dynamics of the universe. In particular, the point $P^\pm_{3v}$ corresponding to the super-accelerated phantom solution, is a late-time attractor and, depending on the chosen initial conditions, it can be the end-point of the cosmic evolution. A phantom solution was discussed in \cite{silva_koyama} for the Brans-Dicke theory with similar cubic derivative interaction $\propto f(\phi)\nabla^2\phi(\der\phi)^2$. However, in this latter case it was a self-accelerating solution and, as shown in section \ref{subsec-no-self-a}, the self-accelerating solution is not compatible with the motion equations \eqref{feqs}, \eqref{kg-eq}, meaning that in the present case this solution does not arise.

So, we are faced with the following problem: there is less asymptotic dynamic structure in the phase space of the model \eqref{action} plus the matter piece action $S_m$, than in the vacuum case (${\cal L}_m=0$ $\Rightarrow$ $S_m=0$), which is a particular case of the former. The answer is in the cosmological equations. In fact, if take a closer look at \eqref{feqs}, \eqref{rho-p}, \eqref{kg-eq}, one can see that the highly non-linear (cubic) derivative interaction $\propto\nabla^2\phi(\der\phi)^2$, boosts the gravitational interactions of the background matter with the galileon, and these screen the gravitational self-interactions of the latter. 

In order to make evident the gravitational interactions of the galileon with the background matter and with itself (what we call as gravitational self-interactions of the galileon), let us write again the Einstein's equations \eqref{feqs}:

\bea &&\;3H^2=\rho_m+\rho_\phi,\nonumber\\
&&-2\dot H=\rho_m+\rho_\phi+p_\phi,\label{feqs-1}\eea where, for simplicity, we are considering pressureless background matter, and the motion equation for the galileon \eqref{kg-eq}:

\bea \ddot\phi+3H\dot\phi-3g_0F=-V_{,\phi},\label{kg-int}\eea where

\bea F=F(H,\dot H,\dot\phi,\ddot\phi)=\left(\dot H+3H^2\right)\dot\phi^2+2H\dot\phi\ddot\phi.\label{f-eq}\eea 

The fact that in the equations \eqref{kg-int}, \eqref{f-eq}, there is not explicit dependence on the energy density of the background matter $\rho_m$, but on the curvature quantities, is a direct manifestation of the minimal coupling between the galileon and the matter fluid. Notice, however, that the galileon indirectly interacts with the background, as well as with itself, through gravity. This is evident in the field equations above. Actually, if one takes into account the Einstein's equations \eqref{feqs-1} and substitutes $H$ and $\dot H+3H^2$ from those equations back into Eq. \eqref{f-eq}, then one gets: 

\begin{widetext}
\bea F=F(\rho_m,\phi,\dot\phi,\ddot\phi)=\frac{\dot\phi^2}{2}\left(\rho_m+2V\right)+\frac{2\dot\phi\ddot\phi}{\sqrt 3}\sqrt{\rho_m+\frac{\dot\phi^2}{2}+V}-g_0\frac{\dot\phi^4}{2}\left(\ddot\phi+\sqrt{3}\,\dot\phi\sqrt{\rho_m+\frac{\dot\phi^2}{2}+V}\right),\label{f-funct}\eea\end{widetext} where we have considered the definitions of $\rho_\phi$ and $p_\phi$ in Eq. \eqref{rho-p}. Now the dependence on the matter density of the background is explicit. If further substitute $F$ from \eqref{f-funct} back into the motion equation of the galileon \eqref{kg-int}, then the gravitational interactions of the galileon with the background matter, as well as the gravitational self-interactions of the galileon, are apparent.

Qualitatively, what happens is that, when we set $\rho_m=0$ in \eqref{f-funct}, the resulting equation of motion \eqref{kg-int}, reflects the existence of the cosmological vacuum solutions which correspond to the critical points in TAB. \ref{tab-2}. These are the result of the gravitational self-interactions of the galileon which are boosted by the cubic derivative interaction through the term $F$ given by Eq. \eqref{f-funct}. Otherwise, if turn-off the cubic interaction by setting $g_0=0$ in Eq. \eqref{kg-int}, only the critical points $P^\pm_{5v}$ and $P^\pm_{6v}$ in TAB. \ref{tab-2}, survive. In such a case ($\rho_m=0$, $g_0=0$), the galileon vacuum is indistinguishable from the quintessence vacuum. Now, if assume non-vanishing background matter ($\rho_m\neq 0$), the gravitational interactions of the galileon with the cosmic background, screen its gravitational self-interactions that led to the existence of the critical points: $P_{1v}$, $P_{2v}$, $P_{3v}$, $P_{4v}$, and $P_{7v}$, which, otherwise, are found in the vacuum galileon case in addition to $P_{5v}$ and $P_{6v}$ (see TAB. \ref{tab-2}).


\section{conclusion}\label{sec-conclu}

The 4D galileon models \cite{nicolis} were inspired by the 5D DGP braneworld \cite{Dvali}. As a matter of fact several of these models share certain similitude with the DGP scenario. In \cite{silva_koyama}, for instance, the Brans-Dicke galileon model given by the following Lagrangian density

\bea {\cal L}_\textsc{sk}=\phi R-\frac{\omega}{\phi}(\der\phi)^2-2\Lambda\phi+f\nabla^2\phi(\der\phi)^2,\label{koyama-lag}\eea was investigated. The authors found a solution that self-accelerates at late-times, which is free of ghost instabilities on small scales. It is required that the BD coupling parameter be a negative quantity ($\omega<0$). At late time gravity is strongly modified and the background cosmology shows a phantom-like behavior. In this paper we have shown that the mentioned similitude between the galileon and the DGP model, in particular the existence of a self-accelerating solution, might not take place.

Here we have investigated the phase space asymptotic dynamics of the generalized galileon model which is depicted by the following Lagrangian density:

\bea {\cal L}=R-(\der\phi)^2-2V_0\,e^{-\lambda\phi}-g_0\nabla^2\phi(\der\phi)^2.\label{model-lag}\eea 

This is a particular case of the model of action \eqref{action}. In the model of \eqref{koyama-lag}, the fourth term in the Lagrangian, i. e. the cubic derivative interaction, is the one which guarantees that small fluctuations around the self-accelerating solution are stable on small scales \cite{silva_koyama}. Meanwhile, in the model \eqref{model-lag}, this term boosts the gravitational interactions of the galileon with the background matter, as well as the gravitational self-interactions of the galileon. The consequence is that, when the background matter is removed, new asymptotic behavior arises with respect to the standard quitessence cosmology (critical points $P_{1v}$, $P_{2v}$, $P_{3v}$, $P_{4v}$, and $P_{7v}$). In the presence of background matter, the gravitational interactions of the latter with the galileon screen the gravitational self-interactions of the galileon vacuum. The result is very interesting and unusual: the asymptotic dynamics of the vacuum of the theory is richer than the asymptotic dynamics of the same model when background matter populates this vacuum.

Our results are unusual enough as to seek for further evidence on the mentioned compensating effects of the background matter on the dynamics of the vacuum of theories of gravity with derivative coupling. In this regard it will be interesting to search for a similar effect in models of the kind in Eq. \eqref{action}, with different functions $g=g(\phi)$ and with other potentials $V(\phi)$ than the exponential. Besides, it could be also interesting to look for similar effects in other Horndeski-type theories with derivative coupling. These ideas will be the subject of future work.


\acknowledgements

TG, UN and IQ want to thank SNI-CONACyT for continuous support of their research activity. The work of RDA was supported by the CONACyT grant No 350411, while GL was supported by Comisi\'on Nacional de Ciencias y Tecnolog\'ia through Proyecto FONDECYT de postdoctorado 2014, grant No 3140244. UN also acknowledges PRODEP and  CIC-UMSNH, while IQ thanks CONACyT of M\'exico for support of the present research project.


\appendix

\section{Poincar\'e projection}

In order to look for possible critical points at infinity, one possibility is to introduce the Poincar\'e variables: $$X_s=\frac{x_s}{r_s},\;Y_s=\frac{y_s}{r_s},\;Z_s=r_s^{-1},\;r_s=\sqrt{1+x_s^2+y_s^2},$$ such that $Z_s=\sqrt{1-X_s^2-Y_s^2}.$ It follows that the dynamical system \eqref{asode-g0-exp} is equivalent to 

\begin{widetext}\begin{align}\label{infinity}
&KX_s'=\sqrt{6}\lambda X_s^2 Y_s^2\left[8 X_s^4+2 X_s^2 \left(8Y_s^2-7\right)+8 Y_s^4-14 Y_s^2+7\right]\nonumber\\
&\;\;\;\;\;\;\;\;\;\;\;\;\;\;\;\;\;\;\;\;\;\;\;\;\;\;\;\;\;\;\;\;\;\;\;\;\;\;\;\;\;\;\;\;\;\;\;\;-6 X_s\left(2Y_s^2-1\right)Z_s \left(4X_s^4+2 X_s^2\left(5Y_s^2-3\right)+6 Y_s^4-9 Y_s^2+2\right),\\
&KY_s'=\sqrt{6}\lambda X_sY_s\left(8 X_s^4\left(Y_s^2-1\right)+2X_s^2\left(8Y_s^4-15Y_s^2+6\right)+8Y_s^6-22Y_s^4+19Y_s^2-5\right)\nonumber\\
&\;\;\;\;\;\;\;\;\;\;\;\;\;\;\;\;\;\;\;\;\;\;\;\;\;\;\;\;\;\;\;\;\;\;\;\;\;\;\;\;\;\;\;\;\;\;\;\;-6Y_s\left(2 Y_s^2-1\right)Z_s\left(4X_s^4+2X_s^2\left(5 Y_s^2-3\right)+6Y_s^4-9Y_s^2+3\right),\end{align}\end{widetext} and the auxiliary equation

\begin{widetext}
\bea K Z_s'=Z_s\left(2 X_s^2+2 Y_s^2-1\right)\left[\sqrt{6}\lambda X_sY_s^2\left(4 X_s^2+4 Y_s^2-5\right)-6\left(2 Y_s^2-1\right)Z_s \left(2 X_s^2+3 Y_s^2\right)\right].\eea\end{widetext} where $$K={2 \left(8X_s^4+20 X_s^2Y_s^2-12X_s^2+12Y_s^4-16Y_s^2+5\right)},$$ can take any sign, and we have rescaled the time derivative by the factor $Z_s.$ By definition $Y_s\geq 0$. Observe that the fixed points at infinity of the system \eqref{asode-g0-exp} are mapped to the fixed point of the system \eqref{infinity} located on the circle $X_s^2+Y_s^2=1.$


In the FIG. \ref{fig-1} are presented phase portraits of the plane-autonomous dynamical system (\ref{asode-g0-exp}) for different values of the parameter $\lambda$. From left to right $\lambda=0$ (constant potential), $1.5$ and $3$ respectively. In the bottom panels the compact (Poincar\'e) phase portrait \eqref{infinity} which corresponds to the whole (finite and infinity) dynamics, in each case, is shown. It is seen that, thanks to the galileon coupling $g_0$, the critical points can be found not only within the semi-disk $x_s^2+y_s^2\leq 1$ (i.e., inside the dashed curve), but also outside it. By numerical inspection we found the fixed points at infinity $Q_1: (X_s=0,Y_s=1)$ and $Q_{2,3}: (X_s=\pm 1,Y_s=0)$.  These are configurations at infinity corresponding to $x_s\rightarrow 0, y_s\rightarrow +\infty$ ($Q_1$) and to  $x_s\rightarrow \mp \infty, y_s\rightarrow 0$ ($Q_2$ or $Q_3$).

In order to determine analytically all the fixed points at infinity one introduces polar coordinates $x_s=r\cos\theta, y_s=r\sin\theta$ and let $\rho=1/r$. Rescaling the time derivative by the factor $\rho$, one obtains 

\begin{widetext}\begin{align}
&\rho'=\frac{1}{4}\rho\left(\rho^2-1\right)\left(\sqrt{6}\lambda\left(5\rho^2+1\right)\sin(\theta)\sin(2\theta)+6\left(\rho^2-5\right)\rho\cos(2\theta)+3\rho\cos(4\theta)-30\rho^3+3\rho\right),\\
&\theta'=\frac{1}{4}\sin(2\theta)\left(\sqrt{6}\lambda\left(-5\rho^4+2\rho^2-1\right)\cos(\theta)+6\left(\rho^3+\rho\right)\left(\cos(2\theta)+\rho^2\right)\right).\end{align}\end{widetext} That is,
	
\bea \rho'=f(\theta)\rho+O\left(\rho^2\right),\;\theta'=g(\theta)+O\left(\rho\right).\eea where 

\bea &&f(\theta)=-\frac{1}{2}\sqrt{\frac{3}{2}}\lambda\sin(\theta)\sin(2\theta),\nonumber\\
&&g(\theta)=-\frac{1}{2}\sqrt{\frac{3}{2}}\lambda\cos(\theta)\sin(2\theta).\nonumber\eea 

The critical points at infinity are found by solving the equations $\rho'=\theta'=0$ on $\rho=0$, that is equivalent to solving $g(\theta )=0$. The solutions $\theta^*$ are given in pairs $\theta_i$ and $\theta_i+2\pi$, however, here we consider only the solutions on the principal interval $0 \leq \theta \leq \pi$. Hence, the angular variable takes values within the interval $0\leq\theta\leq\pi$, since by definition $y_s\geq 0$. The obtained solutions correspond to the following critical points $Q_i:(\theta^i,X_s^i,Y_s^i)$ $$Q_1:(\pi/2,0,1),\;Q_2:(0,1,0),\;Q_3:(\pi,-1,0).$$ There are not other fixed points at infinity. 

We want to notice that one or several of the above critical points may be degenerate. Actually, given the definition of the variables $X_s$ and $Y_s$, and the relationships $x_s=r\cos\theta$, $y_s=r\sin\theta$ and $\rho=1/r$, it is not difficult to see that, under the Poincar\`e projection, those points at infinity for which $x_s$ is finite while $y_s\rightarrow\infty$, are mapped into the single (degenerate) point $Q_1:(\pi/2,0,1)$. In particular, the critical points in TAB. \ref{tab-2}, $P^\pm_{1v}$, for which $x_s=0$, $y_s\rightarrow\infty$, and the phantom critical points $P^\pm_{3v}$, which correspond to the case where $x_s=\mp 2\sqrt{6}/\lambda$, $y_s\rightarrow\infty$, are mapped into the mentioned degenerate point $Q_1$. This reflects the fact that different choices of variables cover just patches of the phase space but not the whole of it, and these should be complemented to get enough information on the dynamics.  

Just for illustration, let us choose the positive branch patch given by the variables $x_+$ and $y$. Using Eq. \eqref{n-var} we get the coordinate relationships:

\bea &&X_s=\frac{(1-x_+)y}{\sqrt{x^2_+y^2+(1-x_+)^2y^2+x^2_+(1-y)^2}},\nonumber\\
&&Y_s=\frac{x_+(1-y)}{\sqrt{x^2_+y^2+(1-x_+)^2y^2+x^2_+(1-y)^2}}.\nonumber\eea This implies that the vacuum critical points $P^+_{1v}:(1,0)$, and $$P^+_{3v}:\left(\frac{\lambda}{\lambda-2\sqrt{6}},0\right),$$ both share the same value of the coordinate $y=0$. Hence, if substitute $y=0$ into the above equations for $X_s$ and $Y_s$, independent of the value of $x_+$, one gets: $$X_s=0,\;Y_s=1.$$ For $P^-_{1v}:(-1,0)$, and $$P^-_{3v}:\left(-\frac{\lambda}{\lambda + 2\sqrt{6}},0\right),$$ we obtain the same result after expressing $X_s$, $Y_s$ in terms of $x_-$, $y$.

\end{document}